\documentclass[journal]{IEEEtran}
\usepackage{cite}
\usepackage{url}

\usepackage{color,soul}

\usepackage{tikz,forest}
\ifCLASSINFOpdf
\usepackage{graphicx}
\usepackage{tabularx}
\usepackage{float}
\graphicspath{{./pdf/}{./jpeg/}}
\else
\fi

\usepackage[cmex10]{amsmath}
\usetikzlibrary{calc}
\usetikzlibrary{positioning}
\usetikzlibrary{decorations.text}
\usetikzlibrary{arrows.meta}
	\usepackage[caption=false]{subfig}

\begin{document}
%
\title{Response Time and Availability Study of RAFT Consensus in Distributed SDN Control Plane }

\author{Ermin~Sakic,~\IEEEmembership{Student Member,~IEEE,}
        and Wolfgang~Kellerer,~\IEEEmembership{Senior Member,~IEEE}%
		\thanks{E. Sakic is with the Department
		of Electrical Engineering and Information Technology, Technical University of Munich, Germany;
		and Siemens AG, Munich, Germany, E-Mail: (ermin.sakic@\{tum.de, siemens.com\}).}%
		\thanks{W. Kellerer is with the Department of Electrical Engineering and Information Technology, 
		Technical University of Munich, Germany, E-Mail: (wolfgang.kellerer@tum.de).}
	\vspace{-9mm}	}
\newcommand{\todo}[1]{\textcolor{red}{#1}\PackageWarning{TODO:}{#1!}}



%


\maketitle

\begin{abstract}
	Software Defined Networking promises unprecedented flexibility and ease of network operations. While flexibility is an important factor when leveraging advantages of a new technology, critical infrastructure networks also have stringent requirements on network robustness and control plane delays. Robustness in the SDN control plane is realized by deploying multiple distributed controllers, formed into clusters for durability and fast-failover purposes. However, the effect of the controller clustering on the total system response time is not well investigated in current literature. Hence, in this work we provide a detailed analytical study of the distributed consensus algorithm RAFT, implemented in OpenDaylight and ONOS SDN controller platforms. In those controllers, RAFT implements the data-store replication, leader election after controller failures and controller state recovery on successful repairs. To evaluate its performance, we introduce a framework for numerical analysis of various SDN cluster organizations w.r.t. their response time and availability metrics. We use Stochastic Activity Networks for modeling the RAFT operations, failure injection and cluster recovery processes, and using real-world experiments, we collect the rate parameters to provide realistic inputs for a representative cluster recovery model. We also show how a fast rejuvenation mechanism for the treatment of failures induced by software errors can minimize the total response time experienced by the controller clients, while guaranteeing a higher system availability in the long-term.
\end{abstract}

\textit{Keywords} - performance analysis, stochastic activity networks, SDN, distributed control plane, RAFT, strong consistency, fault tolerance, smart grid, OpenDaylight, ONOS

%
\IEEEpeerreviewmaketitle

\vspace{-0.4cm}
\section{Introduction}
\subsection{Background and problem statement}
In critical infrastructure, such as the utility \cite{petroulakis2016virtuwind, newsmartgrid} and automotive\cite{sommer2013race} domains, resilience of the communication network is a necessary property and an important criterion for adopting a new and disruptive network technology such as Software Defined Networking (SDN). In single controller SDN scenarios, unavailability of the controller leads to loss of control and monitoring channels with the network devices and hence a system instability. The loss of network control may further result in production and power outages (smart grid \cite{petroulakis2016virtuwind}) or even life-threatening scenarios (dependable automotive \cite{sommer2013race}). 

To address the resilience issues, SDN controllers can be logically coupled into controller \emph{clusters}, where each instance of the controller, referred to as \emph{replica} hereafter, is responsible for managing a number of switches in the network. A particular controller may exhibit its control only over the switches to which it is assigned. In order to provide a fallback solution in case of another controller's failure, it also keeps track of the internal state information related to the switches managed by other controllers.  When a controller replica fails, a different controller instance from the same cluster takes over and resumes operation with some downtime. To keep the backup replicas up-to-date w.r.t. the internal controller state, controllers synchronize their state. Depending on the consistency model which defines the ordering of synchronization messages, the synchronization procedure imposes a varying overhead on the control channel \cite{Sakic, muqaddas2016inter}. The two major controller platforms OpenDaylight \cite{medved2014opendaylight} and ONOS \cite{berde2014onos} implement the \emph{strong consistency model}, which requires that the update of a distributed state has been seen by the majority of the cluster members before it is considered to be committed.

In a strongly consistent cluster, whenever an update request is initialized by a cluster client at one of the controller replicas, the receiving replica sends out the received request to the current cluster \emph{leader}. The \emph{leader} is the controller instance that orders all incoming state update requests, so as to allow for a serialized history of updates and thus operational state consistency at runtime. Following a state update at the leader, the update is propagated using a consensus protocol to the cluster replicas, and is committed to the data-store only after the majority of replicas have \emph{agreed} on the update.

A consensus algorithm ensures that all replicas always decide on the same value (\emph{agreement}), with the constraint that only a value proposed by one of the replicas eventually becomes accepted after the synchronization procedure (\emph{integrity}). Google's Chubby \mbox{\cite{burrows2006chubby}} is a distributed locking service whose state-distribution and failure tolerance mechanism are based on a variation of the Paxos consensus algorithm \mbox{\cite{lamport1998part, moraru2013there}}. OpenDaylight and ONOS implement the more recent algorithm RAFT \mbox{\cite{ongaro2014search}}. Unlike Paxos, RAFT also provides for persistent logging and state reconciliation for recovered replicas.

In addition to the availability concerns, critical infrastructure providers often have very stringent requirements on the experienced control plane delay. For example, the smart grid is a delay-sensitive infrastructure that requires techniques which identify and react on any abnormal communication network changes in a timely manner. If the detection and responses are not made promptly, the grid may become inefficient or even unstable and cause further catastrophic failures in the entire network \cite{newsmartgrid}. Events in the grid may require rapid reaction from the network controller - i.e. rerouting in case of power grid failures, expedited diagnostics and alarm handling \cite{petroulakis2016virtuwind}.
Furthermore, network management systems in the 5G context can require bounded configuration times when establishing on-demand network services \cite{Sakic}. However, the frequently deployed strong consistency model in a clustered SDN requires that, prior to any operation in the SDN controller cluster, a cluster-wide synchronization must occur. The response time of such a control plane is hence dependent on parameters such as the cluster size, controller placement, processing delays in different system components and the failure vectors. 

The clustered SDN controller solutions require estimations of the worst-case response times and the expected availability for arbitrary sets of configurations, before their deployment can be considered in critical infrastructure networks. To our best knowledge, no prior work has investigated these issues. Hence, we fill the gap with an appropriate analytical study. 

In the remainder of the introductory section, we give an overview of our contributions. In Section \ref{background}, we describe the assumed multi-controller SDN architecture from the system perspective. In the same section we specify the technique of Stochastic Activity Networks-based (SAN) modeling and outline the consensus algorithm RAFT. In Section \ref{sanmodels} we introduce and explain in detail the proposed SAN models for response time, controller failure injection and cluster recovery modeling. In Section \ref{evaluationm} we explain the evaluation methods and parametrizations used to compute the results presented in Section \ref{resultss}. In Section {\ref{relatedwork}} we present existing work in the field of distributed SDN control plane and consensus algorithms. Section \ref{conclude} concludes the paper.

\vspace{-0.3cm}
\subsection{Our contribution}

In this paper, we present a system model of a distributed SDN control plane that leverages the Stochastic Activity Networks (SAN) modeling framework for the estimation of cluster response time and availability measures. Our SANs comprise the detailed sub-models of the RAFT consensus algorithm, cluster failure and recovery. We also define the parametrizations (studies) for different cluster configurations in order to evaluate the introduced models. We further evaluate a steady-state configuration of the distributed SDN control plane using long-term failure rates for SDN controllers at sub-module, process and hardware level, and short-term response times experienced after immediate controller failures.
 
By assuming reliable event delivery and bounded network, application and data-store commit delays, we can provide stochastic delay guarantees for response handling times in non-failure, partial-failure and cluster-majority failure states. Failures are modeled as stochastic arrival processes for long-term, and deterministic occurrences for worst-case evaluations. As per the nature of the modeled consensus algorithm RAFT, the recovery process too is a combination of stochastic and deterministic message and timeout delays. We further introduce an enhancement to the current controller platforms for enabling a fast recovery of controller bundles and processes. We evaluate its benefits w.r.t. to the total expected response time and cluster availability using the developed model. Our SAN models are compiled into Continuous Time Markov Chain (CTMC) state spaces. In contrast to existing works on consensus algorithms that derive their performance analysis from experiments, we provide analytical guarantees. To this end, our numerical solutions cover the space of all possible state combinations which an SDN cluster may be in.

\section{System Model, SAN Performability Modeling and the RAFT Consensus Algorithm}
\label{background}
In this section, we introduce the assumed system model that comprises the forwarding devices, multiple SDN controllers for redundancy and controller clients. We then outline the background on the formal concepts used in our modeling and discuss the evaluated RAFT consensus algorithm in more detail. The notation used henceforth is presented in Table I.

\vspace{-0.4cm}
\begin{table}[htb]
	\centering	
	\caption{Notation used in Sections {\ref{background}} and {\ref{sanmodels}}. The remaining model parameters are specified in Table II.}
	\begin{tabular}[htb]{ l l }
		  \hline 
		  Symbol & Parameter \\
		  \hline
		  $C$ &  Number of SDN controller cluster replicas \\
		  $F$ or $N_F$ &  Controller failure count \\
		  $\pi(t)$ & State probability vector at time $t$ \\
		  $q$ &  Uniformization rate constant of the CTMC \\
		  $T_R$ &  RAFT replica-to-leader network delay \\
		  $T_{M}$ &  Actual follower-majority-to-leader network delay \\
		  $T_{M_{worst}} $ &  Worst-case follower-majority-to-leader network delay \\
		  $T_{M_{best}} $ &  Best-case follower-majority-to-leader network delay \\
		  $R_M$ & Number of missing RAFT terms in a \emph{lagging} follower \\
		  $F_{Sf}$ &  Boolean depicting a failure of a single RAFT follower \\
		  $F_{Mj}$ &  Boolean depicting a failure of the RAFT follower majority \\
		  $F_{Ldr}$ &  Boolean depicting a failure of the RAFT leader \\
		  $L_{Up}$ &  Counter of currently available RAFT leaders \\
		  $F_{Up}$ &  Counter of currently available RAFT followers \\
		  \hline
	\end{tabular}
\vspace*{-0.2cm}
	\label{notation}
\end{table}

\vspace{-0.5cm}
\subsection{Generic system model}
We assume a set of \emph{C} SDN controllers, collected in a single cluster and deployed for the purpose of achieving fault-tolerant operation \cite{muqaddas2016inter}. Fig. \ref{fig:systemarch} depicts a deployment of the redundant control- and data planes in an exemplary industrial SDN network. Control plane redundancy is realized by running $C=3$ controllers simultaneously, and a number of disjoint paths in between them for fail-over purposes in case of link and node outages. In general, a deployment of  $C=2*F+1$ controllers tolerates a maximum of $F$ controller failures before the SDN cluster becomes unavailable. Thus, in Fig. \ref{fig:systemarch} only a single controller failure is tolerated before the cluster stops to serve the clients' requests (consult the explanation below). The clients of the controllers, such as the network administrators, switches, network appliances and end-hosts, can trigger controller events that lead to a cluster-wide state synchronization and subsequent event processing in the cluster leader. Clients communicate their requests (i.e. Remote Procedure Calls, state updates, topology events etc.) to any \emph{live} replica that is a member of the SDN cluster. The replica then contacts their current cluster leader to serialize (order) the request, which in return distributes the request to other replicas, commits the request and executes its local state machine (in zero-failure case). The replica is then notified of the result of the request execution and can respond to the client with an application response. In the case of a leader failure during the request processing, a new leader is elected by executing a consensus algorithm and the synchronization process is re-initiated.

The limitation of supporting only $F$ failures when $2*F+1$ replicas are deployed relates to the \emph{CAP theorem} \mbox{\cite{gilbert2002brewer}}. This theorem states that any distributed system can provide a maximum of two of the following three system properties at the same time: consistency, availability and partition-tolerance (CAP). Consensus algorithms such as RAFT \mbox{\cite{ongaro2014search}} and Paxos \mbox{\cite{lamport2001paxos, lamport1998part}} favor consistency and partition-tolerance properties, and are able to forward their state \emph{consistently} even in the face of network partitions. A consistent operation of a controller cluster ensures that the majority of controllers will have the same controller state at any given time, and that no two conflicting state updates are ever successfully committed to the shared update history. Hence, controllers are in \emph{consensus} with regards to their state. Consistent and partition-tolerant operation, however, comes at the cost of a lower availability, since consistent operation in the face of network partitions can only be guaranteed by disabling the operation of a partitioned cluster minority while the majority continues to operate. In the remainder of the paper we take this limitation into account and consider the system as \emph{available} only when the majority of controller nodes are available and are mutually reachable. 
\vspace{-0.8cm}
\newcommand*\ab{.4}
\tikzset{
    net node/.style = {},
    net connect/.style = {line width=1pt, draw=blue!50!cyan!25!black},
 }

\newcommand*\client{\includegraphics[width=.11\textwidth]{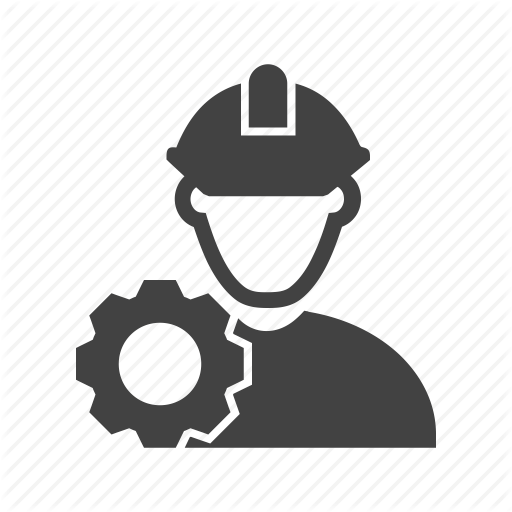}}
\newcommand*\switch{\includegraphics[width=.04\textwidth]{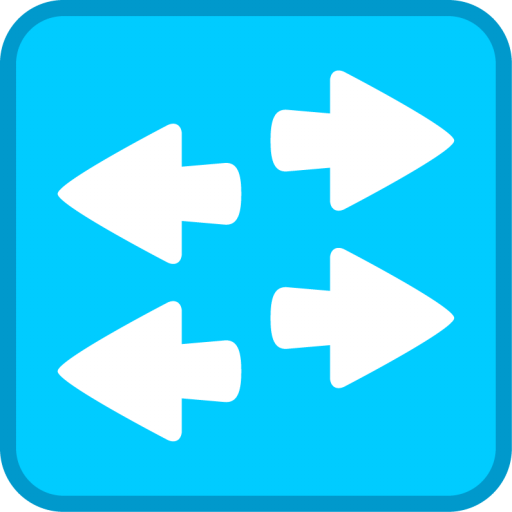}}
\newcommand*\controller{\includegraphics[width=.09\textwidth]{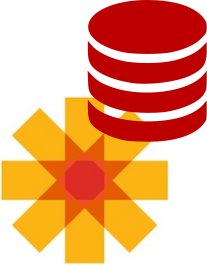}}
 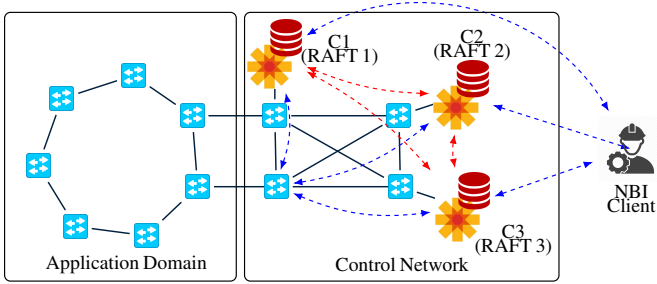
\begin{figure}[htb]
	 \scalebox{.55}{
    \centering
    \begin{tikzpicture}[thick,scale=1, every node/.style={scale=0.8}] 
      \foreach \i in {0,...,6}
	\path (135+\i*51:2) node (n\i) [net node] {\switch};
      \foreach \i in {1,...,6}
	     \pgfmathtruncatemacro{\jn}{\i-1}%
	\path [net connect]
		(n\i)
		-- (n\jn);
      \path [net connect]
	(n0)
	-- (n6); 
      \path [net connect]
	(n5)
	  edge node (n7) [net node, pos=1] {\switch} +(20mm,0mm);
      \path [net connect]
	(n4)
	  edge node (n8) [net node, pos=1] {\switch} +(19mm,0mm);
      \path [net connect]
	(n7)
	  edge node (n9) [net node, pos=1] {\switch} +(30mm,0mm);
      \path [net connect]
	(n8)
	  edge node (n10) [net node, pos=1] {\switch} +(30mm,0mm)
	  -- (n7);
      \path [net connect]
	(n9)
	  -- (n8);
      \path [net connect]
	(n10)
	  -- (n7);
      \path [net connect]
	(n10)
	  -- (n9);
      \path [net connect]
	(n9)
	  edge node (c1) [net node, pos=1] {\controller} +(15mm, 5mm);
      \path [net connect]
	(n10)
	  edge node (c2) [net node, pos=1] {\controller} +(15mm, -5mm);
      \path [net connect, Latex-Latex, dashed, color=blue, thick]
	(c1)
	  edge node (cl1) [net node, pos=1] {\client} +(41mm, -13mm);
      \path [net connect, Latex-Latex, dashed, color=blue, thick]
	(c2)
	-- (cl1); 
     \path [net connect]
	    (n7)
		edge node (c3) [net node, pos=1] {\controller} +(0mm, 15mm);
	    -- (n7);
      
      \draw[bend right=-45, Latex-Latex, color=blue, thick, dashed] (c3) to (cl1);
      \draw[bend left=-10,Latex-Latex, color=red, thick, dashed] (c3) to (c1);
      \draw[bend left=-10,Latex-Latex, color=red, thick, dashed] (c2) to (c3);
      \draw[bend left=-10,Latex-Latex, color=red, thick, dashed] (c1) to (c2);
      \draw[bend left=-15,Latex-Latex, color=blue, thick, dashed] (n8) to (c1);
      \draw[bend left=-15,Latex-Latex, color=blue, thick, dashed] (n8) to (c2);
      \draw[bend left=-15,Latex-Latex, color=blue, thick, dashed] (n8) to (c3);

      \draw[black,thick,rounded corners] (-2.8,-3) rectangle (2.8,3.5);
      \draw[black,thick,rounded corners] (3,-3) rectangle (10.6,3.5);

      \node[text width=7cm] at (1,-2.6) {\LARGE Application Domain};
      \node[text width=7cm] at (8.0,-2.6) {\LARGE Control Network};
      \node[text width=3cm, align=center] at (5.3,2.6) {\LARGE C1 (RAFT 1)};
      \node[text width=7cm, align=center] at (8.5,2.7) {\LARGE C2\\(RAFT 2)};
      \node[text width=7cm, align=center] at (9.5,-2.0) {\LARGE C3\\(RAFT 3)};
      \node[text width=7cm, align=center] at (12.35,-1) {\LARGE NBI\\Client};
    \end{tikzpicture}
    }
    \caption{An exemplary industrial SDN with redundant paths for majority of controller-to-controller and controller-to-switch connections.  The SDN controllers execute the RAFT \cite{Howard:2015:RRW:2723872.2723876} agents, responsible for per-state-shard state synchronization, leader election and cluster recovery after individual replica failures. Red dashed lines represent the RAFT session exchanges between the SDN controller replicas, blue dashed lines are the "client" connections (switch-controllers, northbound interface client-controllers).}
\vspace*{-0.2cm}
    \label{fig:systemarch}
  \end{figure}

In state of the art controller platform implementations with a strong consistency model \cite{medved2014opendaylight,berde2014onos}, network configuration requests facilitate a number of state changes and inter-controller synchronization steps before coming to a consensus in decision and actual execution of the configuration change. 
For example, assuming an SDN module that subscribes and reacts to topology changes (e.g., raises alarms to an administrator in case of link failures), the topology change would first need to be committed across the majority of controllers, before the subscribed SDN module could be notified of the committed change and execute the reaction. The duration of this process obviously depends on the cluster size, the availability of controllers and the controller-to-controller delays.

In the following subsections we describe the SAN framework we use for modeling of the distributed SDN control plane. We give an overview of the RAFT consensus algorithm responsible for state synchronization, cluster leader election and state recovery after failures. 

\vspace{-0.4cm}
\subsection{Stochastic Activity Networks}
	In contrast to previously eublished methods on the evaluation that provides a mean to evaluate an \emph{existing} product or deployment, e.g. by using measurement techniques, \emph{deductive} analysis allows for a system evaluation \emph{before} the system is actually deployed. Hence, significant savings can be achieved if the deductive solutions are able to accurately predict the real-world behavior of the future non-implemented system or system extensions. With this in mind, contrary to the previously published methods on evaluation of the distributed SDN control plane, which base their analysis on a limited number of physical cluster configurations \mbox{\cite{suh2017toward, suh2016performance}}, we opt for the flexible and economical deductive solution.

\emph{Discrete-event simulation} is, for example, partially applicable to our problem. Simulation allows for a tunable quality of the results by repeating execution of a given model and derivation of the relevant output measures. However, the simulation methodology may not handle corner cases, which are numerous in a consensus algorithm such as RAFT.

Another class of deductive analysis methods are the \emph{analytic numerical} methods, which are suitable when a closed-form solution is not obtainable. Analytic numerical solvers allow for an accurate evaluation of each system state configuration. For this purpose, they require a manually or automatically generated model state space as an input. The additional overhead of the state space generation, as well as the inclusion of each state in the solution, generally leads to a higher computational effort compared to simulation. Furthermore, the generation of the state space may lead to a state explosion problem and infeasible solving times. Therefore, we dedicate Section {\ref{ssec:scalability}} to specifically discuss the scalability of our models. Instead of the manual state modeling, we automate the model generation process and hence avoid the issue of \emph{largeness} \mbox{\cite{trivedibook}} of the resulting state space. For the purpose of the automated model generation, we use the Stochastic Activity Networks (SANs), one of the most prominent representatives of model generation frameworks. We choose specifically SANs over similar techniques, such as Generalized Stochastic Petri Nets and Stochastic Reward Nets because of its practical extensions for the \emph{inhibition} of state transitions, as well as the flexible predicate assignment to the \emph{gate} abstractions (see below).

SANs are an extension of Petri Nets (PN) and an established graphical language for describing the system behavior. SANs have been successfully used in survivability and performability studies of critical infrastructures \cite{avritzer2015survivability}, industrial control systems \cite{ndiaye2016performance} and telecommunication systems \cite{gonzalez2015service} since the late 1980s. We provide a brief summary of the most important SAN concepts used in our modeling here, and refer the more interested reader to comprehensive descriptions in \cite{trivedibook, sandersformalisms}. 

	A SAN consists of \emph{places}, \emph{activities}, \emph{input gates} and \emph{output gates}. Similar to PNs, places have a certain \emph{token} assignment associated with them. Every unique assignment of tokens across the places uniquely defines a state of the SAN. These states are called \emph{markings}. In Markov Chain analogy, a single marking represents a unique state of a Markov Chain. An \emph{activity} element of a SAN defines a transition with the corresponding transition rates, and allows for controlling the flow of tokens from a single SAN place into a different SAN place. Furthermore, an activity allows for connecting a place to an \emph{output gate} where, on transition of a token from a place to an output gate, a sequence of actions can be taken - e.g. "\emph{if the number of tokens in place A $\textgreater$ n $\rightarrow$ increment the number of tokens in place B by $m$}". Hence, compact state changes (and a large number of unique markings) triggered by a particular transition may be modeled using a smaller number of modeling elements, compared to a traditional Markov Chain.

When an activity \emph{fires}, a number of tokens are removed from the source place and transferred to a destination place connected by the activity. An \emph{input gate} serves as an inhibitor of an associated activity. It specifies a boolean predicate which, when evaluated \emph{true}, enables an activity and allows the firing of the activity. If the inhibitor evaluates \emph{false}, the associated activity is disabled. An \emph{instantaneous} activity is enabled at all times, and will fire whenever there are tokens available in its input place. A \emph{timed} activity, on the other hand, is assigned a \emph{time distribution function} which specifies the firing rate of a specific activity. In our model, for timed activities we assign the deterministic (Erlang-approximated) and exponential firing rates, but also specify instantaneous activities where necessary. An activity may further lead to a token transfer from a source place to one of multiple destination places. This uncertainty is modeled using a \emph{case} definition for each destination state, where each case is assigned a probability parameter.

	To solve the SAN, it must first be transformed into a discrete-state stochastic process \cite{trivedibook}. We make use of the flat state space generator implemented in the M\"obius modeling tool\cite{mobius}, to generate the Continuous Time Markov Chain (CTMC) state space inherent to the evaluated SAN. 
To derive instantaneous state probabilities of a CTMC, the transient solver of M\"obius implements the \emph{uniformization} method \cite{reibman1988numerical, trivedibook}. In short, using uniformization, the \emph{transient state probability vector} $\pi(t)$ of the CTMC can be expressed in terms of a one-step probability matrix of a Discrete Time Markov Chain (DTMC), so that all state transitions of a resulting DTMC occur with a \emph{uniform} rate $q$. As a result of the transformation, the desired state probability vector $\pi(t)$ at time $t$ is governed by a Poisson variable $qt$ and can be expressed as follows:
		\begin{equation}
			\pi(t) \approx \sum_{i=l}^{i=r}{v(i) e^{-qt} \frac{(qt)^{i}}{i!}}\ \mathrm{where}\ v(0) = \pi(0)  
		\end{equation}
where $v(i)$ represents an iteratively computed DTMC state probability vector at step $i$. Lower and upper bounds, $l$ and $r$, govern the number of iterations required to compute the state probability vector with an overall error tolerance of $\varepsilon=\varepsilon_l+\varepsilon_r$ and truncation points $l$ and $r$, respectively.
\vspace{-0.4cm}
\subsection{Case Study: RAFT Consensus Algorithm}
\label{ssec:raft}
	RAFT is a distributed consensus algorithm that provides safe and ordered updates in a system comprised of multiple running replicas. RAFT is the only consensus algorithm implementation in the two prominent open-source SDN controller platforms OpenDaylight\cite{medved2014opendaylight} and ONOS\cite{berde2014onos}. It tries to solve the issues of understandability of the previous de-facto standard consensus algorithm Multi-Paxos \cite{lamport1998part}, and additionally standardizes an implementation of leader election and post-failure replica recovery operations.  A comprehensive description of the algorithm can be found in \cite{ongaro2014search, Howard:2015:RRW:2723872.2723876}. 

\tikzset{slow node/.style={circle,fill=orange!20,draw, text width=2cm, minimum size=2cm, inner sep=0.3pt},}
\tikzset{down node/.style={circle,fill=red!20,draw, text width=2cm, minimum size=2cm,inner sep=0.3pt},}
\tikzset{main node/.style={circle,fill=green!20,draw, text width=2cm, minimum size=2cm,inner sep=0.3pt},}
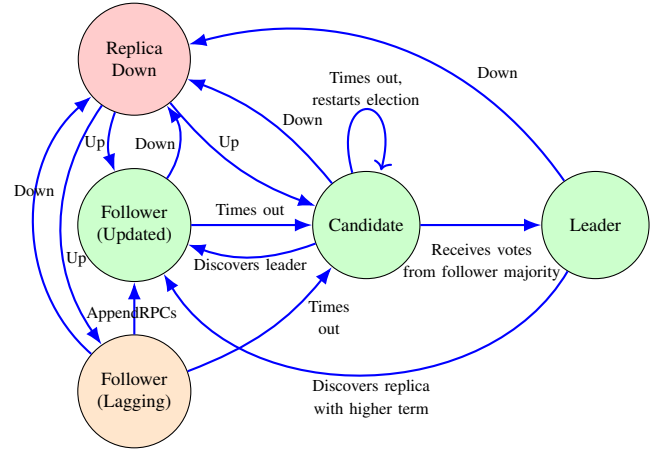
\begin{figure}[htb]
	\centering
		\centering
		\hspace*{-1cm}\begin{tikzpicture}[scale=0.7, transform shape]
			\node[main node, align=center] (1) {Follower \\ (Updated)};
			\node[down node, align=center] (0) [above = 1cm of 1] {Replica \\ Down};
			\node[slow node, align=center] (4) [below = 1cm of 1] {Follower \\ (Lagging)};
			\node[main node, align=center] (2) [right = 2.3cm of 1] {Candidate};
			\node[main node, align=center] (3) [right = 2.3cm of 2] {Leader};

			\draw[bend left=0,-Latex, color=blue, thick] (1) to (2);
			\draw[bend left=20,-Latex, color=blue, thick] (2) to (1);
			\draw[bend left=59,-Latex, color=blue, thick] (3) to (1);
			\draw[bend left=0,-Latex, color=blue, thick] (2) to (3);
			\draw[bend left=-15,-Latex, color=blue, thick] (0) to (2);
			\draw[bend left=-35,-Latex, color=blue, thick] (0) to (4);
			\draw[bend left=-20,-Latex, color=blue, thick] (0) to (1);
			\draw[bend left=-35,-Latex, color=blue, thick] (1) to (0);
			\draw[bend left=-15,-Latex, color=blue, thick] (2) to (0);
			\draw[bend left=-35,-Latex, color=blue, thick] (3) to (0);
			\draw[-Latex, color=blue, thick] (4) to (1);
			\draw[bend left=49, -Latex, color=blue, thick] (4) to (0);
			\draw[bend right=15, -Latex, color=blue, thick] (4) to (2);
		
			\path[->] 
				(2) edge [loop above, -Latex, color=blue, thick] node {} ();
			
			\node[text width=3cm, align=center] at (4.4, 2.6) {\small Times out, \\ restarts election};
			
			\node[text width=3cm, align=center] at (4.5,-3.3) {\small Discovers replica \\ with higher term};
			\node[text width=3cm, align=center] at (6.6,-0.7) {\small Receives votes\\from follower majority};
			\node[text width=3cm, align=center] at (2.2,0.3) {\small Times out};
			\node[text width=3cm, align=center] at (2.2,-0.75) {\small Discovers leader};
			\node[text width=3cm, align=center] at (6.9,2.9) {\small Down};
			\node[text width=3cm, align=center] at (3.2,2.0) {\small Down};
			\node[text width=3cm, align=center] at (.4, 1.55) {\small Down};
			\node[text width=3cm, align=center] at (-0.75,1.55) {\small Up};
			\node[text width=3cm, align=center] at (-1.1,-0.65) {\small Up};
			\node[text width=3cm, align=center] at (-1.9,0.65) {\small Down};
			\node[text width=3cm, align=center] at (1.8,1.55) {\small Up};
			\node[text width=3cm, align=center] at (-0.05,-1.7) {\small AppendRPCs};
			\node[text width=3cm, align=center] at (3.7,-1.75) {\small Times \\ out};
				\end{tikzpicture}

		\caption{A simplified lifecycle schema of a replica inside the RAFT cluster. Adapted from \cite{Howard:2015:RRW:2723872.2723876} and extended for the purpose of detailed modeling.}
\vspace*{-0.3cm}
		      \label{fig:raftlifecycle}
\end{figure}

	A RAFT cluster comprises \emph{leader}, \emph{follower} and \emph{candidate} replica roles. The leader is the node that parses and distributes incoming client updates (i.e. \emph{reads}, \emph{writes}, \emph{no-ops}) to RAFT followers and ensures safe commits. The majority of cluster followers must confirm the acceptance of a new update before the leader and the followers may commit the update in the local \emph{commit log}. Only after the update is committed, the SDN applications built on top of a RAFT agent can continue their processing. After the application has computed the operation related to state update, a response is forwarded to a client (e.g., a switch or network management system). RAFT guarantees that the applied state updates are eventually committed in every available replica in the cluster in the right order. Furthermore, each update is applied \emph{exactly once}, hence enabling \emph{linearizable semantics} \cite{Howard:2015:RRW:2723872.2723876} when operating with the controller state. In the case of a leader failure, after an expiration of an internal follower timeout, the remaining followers automatically switch to a candidate role. A \emph{candidate} is an active replica which offers to become the new cluster leader. To do so, it propagates its candidate status to the other available replicas. If a majority of nodes vote for the same candidate, this candidate node becomes the new leader.

	Updates in RAFT require a single-round trip delay between the leader and the \emph{preferred follower majority} (the fastest to reach followers). When a controller failure occurs, depending on the role of the failed replica, additional delay overhead is imposed. Failures in the RAFT leader during the processing of a particular update lead to a new leader election after an expired \emph{election timeout}. After an exceeded \emph{client timeout}, the client retries its request. If instead of the leader a follower had failed, depending on the follower's type \emph{and} the number of active followers, we distinguish three scenarios:
	{\begin{itemize}
		\item Failure of a follower that \emph{is not a member} of the preferred follower majority results in no additional imposed delays between the leader and the cluster majority.
		\item Failure of a follower that \emph{is a member} of the preferred follower majority leads to the RAFT leader having to include an additional "slower" follower in the preferred follower majority. This, in return, may negatively affect the update commit times depending on the follower's placement and its distance to the RAFT leader.
		\item Failure of any follower that comprises the follower set, with no backup followers available (stand-by RAFT members), necessarily leads to the cluster unavailability. The client update requests that were not successfully committed must be repeated by the client.
	\end{itemize}

		After a successful recovery of the majority of the RAFT members and the re-election of a new leader, RAFT is able to forward its state and commit new updates. Depending on the failure source and the repair time, as well as on the RAFT recovery parameters (\emph{candidate} and \emph{election timeout}), the recovery takes a non-deterministic period to finish.
	
	Fig. \ref{fig:raftlifecycle} gives a high-level overview of the states a cluster replica may traverse throughout its lifecycle. We present the more detailed structural and behavioral models of RAFT in zero- and multiple-failure cases in Section \ref{sanmodels}. Section \ref{evaluationm} details the timing variables used in our parametrization of RAFT.

 \begin{figure*}[htb]
		\centering
		\includegraphics[width=1.05\textwidth]{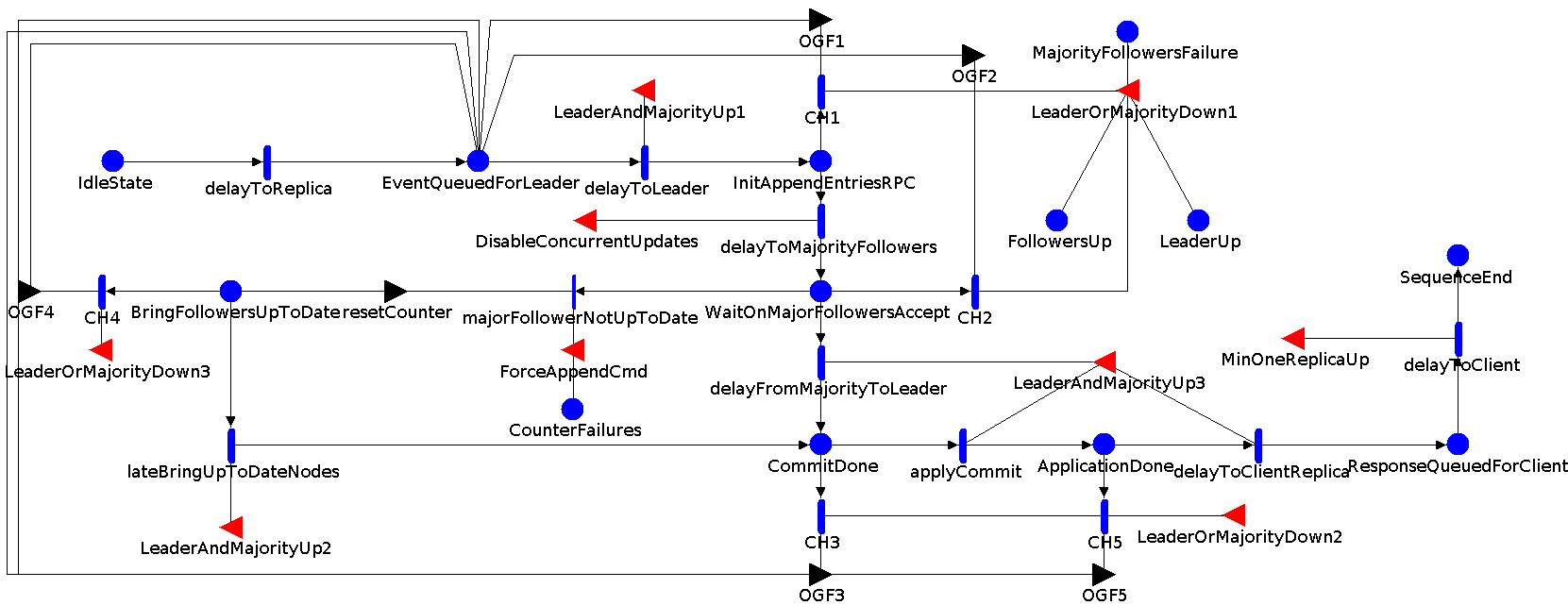}
		\caption{The RAFT response time model depicting the sub-processes of: the receival of a client event at a follower proxy-replica; the event propagation to the RAFT leader; the event propagation from the leader to the follower majority; the data-store commit of the client event and its subsequent processing in the SDN application; and ultimately the propagation of the SDN application's response from leader, through the proxy-replica, to client. Detecting a failure of the RAFT follower majority or leader leads to the restart of the event handling process, starting at the place \texttt{EventQueuedForLeader} - but only after an added deterministic delay of \emph{client timeout} (see Table \ref{tab:generalmodel}). Furthermore, the extended sub-process of \emph{AppendEntries RPC}, necessary to reconcile the RAFT followers that lag behind the current RAFT leader in terms of their state, is included in the lower right part of the critical path (state \texttt{BringFollowersUpToDate}).}
\vspace*{-0.3cm}
		\label{fig:e2emodel}
 \end{figure*}

\vspace{-0.3cm}
\section{SAN Models}
\label{sanmodels}
In this section, we present the SAN models for response time, failure and recovery processes in the context of a RAFT-enabled SDN control plane. We represent \emph{places} as blue circles, \emph{timed activities} as thick blue vertical bars, \emph{instantaneous activities} as thin blue vertical bars, and \emph{input} and \emph{output gates} as thick red and black arrows, respectively.

\vspace{-0.3cm}
\subsection{RAFT End-to-End Delay SAN Model}
\label{ssec:e2emodel}
\def\lc{\left\lceil}   
\def\rc{\right\rceil}
\def\lf{\left\lfloor}   
\def\rf{\right\rfloor}

The distributed SDN control plane model assumes \emph{C} controllers connected in a RAFT cluster, hosting one or multiple SDN applications (referred to as \emph{bundles}) that react to asynchronous \emph{client} events. The \emph{client} device is external to the SDN controller (e.g. an OpenFlow switch or a northbound interface consumer). The client sporadically generates events, such as flow requests or switch notifications which necessitate data-store updates and its subsequent synchronization. The client delivers these events in asynchronous and reliable manner to the replicas in the cluster for processing. After the controller cluster finishes the event processing, the client is notified of the result. The SAN in Fig. \ref{fig:e2emodel} depicts this process. 
 
The place \texttt{IdleState} models the initial system state where no events are queued for internal processing. Following a new event arrival at \emph{any} of the RAFT replicas, the receiving replica is tasked with the propagation of the new event to the current RAFT leader. New event arrivals increment the token amount in the state \texttt{EventQueuedForLeader}, where events are queued until a leader controller replica becomes elected in the cluster. The input gates enumerated \texttt{LeaderAndMajorityUp\#} ensure that the transmission of the event to the RAFT leader or replicas, as well as the intermediate processing inside the cluster happens only in the case where both the RAFT leader and the follower majority are up and available in the cluster. Propagation of the event from the furthest-away replica to the leader is modeled by the activity \texttt{delayToLeader} using a deterministic worst-case delay metric $T_R = T_{M_{worst}}$ (see below). On a received event, the leader initiates the propagation of the respective data-store state update to its followers. The delays induced by the activities \texttt{delayToMajorityFollowers} and \texttt{delayFromMajorityToLeader} correspond, in the best-case to the leader-to-(preferred)-majority delay $T_{M_{best}}$. In the worst-case, to contact the follower majority, the leader needs to contact the follower furthest away from it and hence induce the worst-case uni-directional network delay $T_{M_{worst}}$. Thus, the delay between the RAFT leader and the follower majority is governed by the number of failed followers. We model the non-constant leader-follower majority delay $T_M$ as detailed below. When the follower majority has acknowledged the state update, the leader continues committing the data-store change locally, and the system eventually reaches the \texttt{CommitDone} state. To prevent the leader from broadcasting multiple unacknowledged updates, we ensure the input gate \texttt{DisableConcurrentUpdates} enables the transition \texttt{delayToMajorityFollowers} if and only if the distribution states of RAFT do not contain any outstanding tokens (no synchronization in progress). 

Alternatively, in the case of at least one replica that necessarily comprises the cluster majority lags behind the RAFT leader in terms of its \emph{commit log} (see Subsection \ref{ssec:raft}), the leader enforces additional steps in order to synchronize the cluster majority with its local view. To this end, the activity \texttt{majorFollowerNotUpToDate} fires and a token is incremented in the place \texttt{BringFollowersUpToDate}. The lagging followers are thus assigned the \texttt{Follower (Lagging)} node status (depicted in Fig. {\ref{fig:raftlifecycle}}). For each RAFT term state that is missing in the lagging follower, an additional round-trip delay in the critical path between the leader and replica is induced, hence adding $2\cdot{R_M}*T_{M}$ to the overall delay, where $R_M$ is the maximum number of missing RAFT terms in the follower. This delay is imposed in the definition of the activity \texttt{lateBringUpToDateNodes}. To govern the activation of the instantaneous activity \texttt{majorFollowerNotUpToDate} we make use of a stateful counter \texttt{CounterFailures} that is incremented on each new logged replica failure (refer to Subsection {\ref{ssec:clusfail}}). We consider the worst-case and hence assume that at \emph{any time} after $\lf (C-1)/2 \rf +1$ replica nodes have been disabled, out-of-date replicas are automatically present in the majority of the follower nodes required to confirm a leader update. Hence, we infer the additional overhead of updating the lagging nodes in the replica majority. The flag to enable the activity \texttt{majorFollowerNotUpToDate} is cleared after the reconciliation (as a side configuration of the gate \texttt{resetCounter}).

Following an applied data-store commit in the majority of replicas, the leader commits the state locally and the SDN application gets notified of the data-store event. The data-store commit and the SDN application's processing delay are induced during the activity \texttt{applyCommit} and are modeled as $T_C$ and $T_A$ in Table \ref{tab:generalmodel}, respectively. After the application has completed its processing (system state \texttt{ApplicationDone}), the leader notifies the replica that initially generated the update event (thus adding once-more $T_R$ to the overall worst-case delay), and the replica further forwards its response to the client (thus adding ${T_{CR}}$ which is the client-replica delay). The system then finally reaches the stable state \texttt{SequenceEnd}, where the event is marked as successfully processed. In case of a failure occurrence in the leader or follower majority during the event processing, the activities named \texttt{CH\#} lead to a token being shifted from the current SAN place to the \texttt{EventQueuedForLeader} place, using the output gate increment action modeled by \texttt{OGF\#}. Hence, the event distribution procedure restarts as soon as the cluster is re-established. The delay until a critical failure occurrence of the leader or the follower majority is noticed by the client is modeled using the \emph{client timeout} $T_{CL}$.

As previously noted, the delay from the RAFT leader to the furthest-away replica from the follower majority will vary depending on the availability and proximity of followers w.r.t. the current cluster leader. We annotate the leader-majority followers delay as $T_M$. Assuming a deployment of $C$ controller nodes and a single leader $L$ at any time, the set $S_L = \{D_{R1}, D_{R2}...D_{R_{\lf C/2 \rf}}\}$ contains the maximum bounded delays between the leader $L$ and $\lf C/2 \rf$ follower nodes \emph{closest} to $L$ w.r.t. delay between controller $L$ and each of the \emph{available} followers $R_I$. Hence, we define the delay between leader $L$ and the follower majority as the delay between $L$ and the \emph{farthest} follower in the majority $T_M = max\{S_L\}$. 

To emphasize the effect of a failed preferred-follower controller on the response time, in our exemplary evaluation, we scale the delay value to contact the followers majority linearly with the number of currently available followers using a scaling factor $S_F$ so that:

\begin{equation*}
	T_M = \begin{cases}
		S_F * T_{M_{best}} & \mathrm{when}\ F_{up}\geq \lf C/2 \rf  \\
	     undefined & \mathrm{otherwise}
\end{cases}
\end{equation*}

For the evaluation purposes we model the $S_F$ as a function of the current marking of SAN so that $S_F=\frac{C-1}{F_{up}}$ and thus:

\begin{equation*}
	T_M = \begin{cases}
		\frac{C-1}{F_{up}} * T_{M_{best}} & \mathrm{when}\ F_{up}\geq \lf C/2 \rf  \\
		undefined & \mathrm{otherwise}
	\end{cases}
\end{equation*}
where $F_{up}$ represents the number of currently available followers. In the best case where all nodes are up, the leader-majority delay equals $T_{M_{best}}$. In the worst case, the controller-majority delay peaks at $T_{M_{worst}} = T_R = 2*T_{M_{best}}$ when only $\lf C/2 \rf +1$ controller nodes (including the leader) are active. 

Using a fixed scaling factor is an exemplary and non-optimal representation, as the exact worst-case leader-majority delay is equal to the delay between the leader and the \emph{farthest} away follower in the current follower majority, and hence necessitates knowing the exact bounded delays between each two SDN controllers in the network. We omit this level of model granularity as the required parameters would require population from an engineered network topology, and would further rely on the optimality of the used controller placement technique. Nevertheless, the SAN model proposed here can be extended to take an arbitrary set of controller-to-controller delay parameters with little effort. 

\emph{Data-store sharding}: The data-store of an SDN controller (e.g., OpenDaylight) is sharded into an arbitrary amount of \emph{data shards} at a flexible granularity (e.g., data shard for topology or flow state). Separate RAFT sessions are responsible for each data shard. We assume that all data shards are available on all SDN controller replicas. Hence, each available controller is an active member of each per-shard RAFT session. RAFT can handle the updates of different data shards concurrently and in isolation. This enhances the overall throughput of the system as multiple asynchronous updates to different shards are parallelized and executed in a non-blocking manner.

\emph{Batching of the data-store updates and latency considerations:} We assume that the clients specify their updates that modify a shard either as a single state update or as a batch of updates \cite{ongaro2014consensus} for maximized throughput and minimized response time. Thus, the worst-case occurs when the update or a batch of concurrent updates is exchanged in a single frame across the cluster and the majority of the cluster members fail before the updates are committed successfully. If a new update arrives during the processing of another update of the same shard \emph{and} a leader fails, we assume that the client updates are in the worst-case batched with the previous non-committed updates and are transferred in one round after the cluster has recovered. This model is fitting for handling real-time events (e.g., alarms) that should preferably never get queued. 

\vspace{-0.4cm}
\subsection{Cluster Failure SAN Model}

\label{ssec:clusfail}
To evaluate the performance of the SDN distributed control plane and RAFT in the face of failures, we introduce a dedicated failure model. For general long-term considerations, we distinguish between hardware and software failures with failure rates $\lambda_{F_H}$ and $\lambda_{F_S}$, respectively. All specified non-deterministic timeouts, failure and repair rates in our model follow a negative exponential distribution. For software failures, we distinguish failures at the application bundle (i.e. an OSGI bundle in ONOS \cite{berde2014onos} and OpenDaylight \cite{medved2014opendaylight} controllers) and process level failures. Similarly, repair rates are distinguished correspondingly as specified in Table \ref{tab:generalmodel}. 

The SAN failure model is depicted in Fig. \ref{fig:failureinjection}. The place \texttt{NodesUp} contains the total number of available controllers (nodes that are up, but not necessarily assigned a RAFT member role). Depending on the failure type (at hardware, process or bundle level), after an occurrence of a failure, a token is placed into the respective \texttt{NodesDown} place. Furthermore, each firing of a failure activity triggers a token addition in the place \texttt{NodeDownSelectFailure} and results in a subsequent evaluation in the instantaneous \emph{case} activity \texttt{failureSelectRole}. We distinguish between the safe-follower ({$F_{Sf}$), follower-majority ($F_{Mj}$) and leader failures ($F_{Ldr}$), with the following probabilities:
	\begin{small}
	\begin{equation*}
		P(F_{Sf}) = \begin{cases} 
			\frac{F_{up}}{F_{up}+1} & \mathrm{when}\ F_{up} \geq \lc (C-1)/2 + 1 \rc \land L_{up} > 0 \\
			 0			& \mathrm{otherwise}
		    \end{cases}
	\end{equation*}
	\begin{equation*}
		P(F_{Mj}) =  \begin{cases} 
			 1 & \mathrm{when}\ F_{up} < \lc (C-1)/2 + 1 \rc \lor L_{up} == 0 \\ 
			 0 & \mathrm{otherwise} 
		\end{cases}
	\end{equation*}
	\begin{equation*}
		P(F_{Ldr}) = \begin{cases} 
			\frac{1}{F_{up}+1} & \mathrm{when}\ F_{up} \geq \lc (C-1)/2 + 1 \rc \land L_{up} > 0 \\
			0 & \mathrm{otherwise}
	        \end{cases}
	\end{equation*}
	\end{small}
where $F_{up}$ and $L_{up}$ are the counters of tokens in places \texttt{FollowersUp} and \texttt{LeaderUp} before the failure occurrence, respectively. Failure of the follower-majority $F_{Mj}$, or a leader failure $F_{Ldr}$ during the event handling in controller, results in a client timeout and subsequent restart of the event handling process. On the other hand, failure $F_{Sf}$ does not affect the cluster availability as the reorganization of a stable cluster majority is still possible with the remaining nodes, albeit with an added delay (as per the definition of $T_M$).

\begin{figure}[htb]
		\centering
		\includegraphics[width=0.5\textwidth]{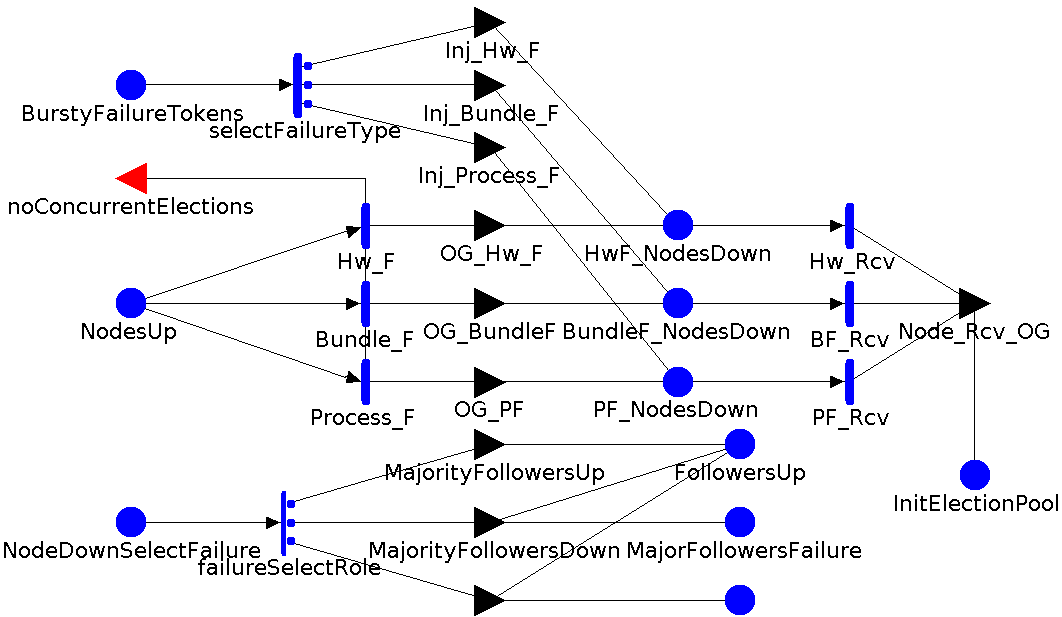}
		\caption{The SAN model of the failure processes includes the long-term failure rates (\texttt{Hw\_F, Process\_F} and \texttt{Bundle\_F}) and the controlled failure injection (\texttt{Inj\_Hw\_F, Inj\_Process\_F} and \texttt{Inj\_Bundle\_F}). The failure type is decided based on a random selection process (bottom-left), and its severity is a function of the current system state (bottom-right).}
\vspace*{-0.3cm}
		\label{fig:failureinjection}
 \end{figure}

In order to observe the response time during and shortly after the failure, we also model a procedure for \emph{controlled failure injection} of single and multiple-correlated transient controller failures and observe the system performance over a short-term time range at millisecond granularity. The correlated failures are modeled as bursty and may occur concurrently. In the past, correlated failures have been investigated in the context of distributed systems \cite{nagaraja2003using}, and represent a flexible method to consider chained failure propagation, i.e. resulting from a malfunctioning replicated SDN application. The failure injection process is depicted in the upper left part of the SAN shown in Fig. \ref{fig:failureinjection}. The place \texttt{BurstyFailureTokens} initially holds a number of tokens corresponding to the number of simultaneous bursty failure injections. The activity \texttt{selectFailureType} governs the probability distributions for encountering a particular type of node failure.

As will be shown in the Subsection \ref{ssec:resultsresp}, in our response time evaluation we distinguish the scenarios of mixed hardware and software failure injection, as well as the single and multiple-correlated failure injections at the granular level of controller bundle, controller process or hardware.

\emph{Critical data plane failures:} In a DTMC, the probability of occurrence of an SDN controller element failure $F_c$ or a critical data plane element failure $F_d$ corresponds to:

\begin{equation}	
P(F_c \cap  F_d) = P(F_c) + P(F_d) - P(F_c \cup F_d)
\end{equation}
}

	In the continuous time domain, failure arrivals for the critical data plane elements that carry the controller-to-controller flows, and failure arrivals for the SDN controller elements can be represented as two independent Poisson processes $N_d(t)$ and $N_c(t)$ with the unique firing rates $\lambda_d$ and $\lambda_c$, respectively. Since the two processes are independent, they also have independent increments. Therefore, critical failure arrivals associated with the summed process $N_t(t) = N_d(t) + N_c(t)$ can be modeled using the rate $\lambda_t = \lambda_d + \lambda_c$.

The failure rates for the critical data plane paths which carry the network control flows can be embedded in the parametrization of our models without an additional modeling overhead (see Table {\ref{tab:generalmodel}}). However, in this work we primarily focus on studying the control plane consensus for the use case of a highly redundant industrial network \mbox{\cite{molina2015availability, molina2016performance}}. Thus, we intentionally decouple our work from the data plane reliability studies and assume the reliable parametrization $1/\lambda_d=\infty$.

\vspace{-0.3cm}
\subsection{RAFT Recovery SAN Model}

The RAFT recovery SAN model in Fig. \ref{fig:raftrecovery} depicts the process of re-inclusion of a previously disabled controller replica in the RAFT cluster. The place \texttt{InitElectionPool} holds a token for each running controller replica that is available but still needs to be admitted in the cluster. As per RAFT design, the replica expects the RAFT leader of the current term to announce its presence using a leader heartbeat. If a leader is identified before the \emph{follower timeout} expires, the replica takes upon the follower role and a token is assigned to the place \texttt{AnnounceFollowerRole}. Alternatively, the replica switches to the \emph{candidate} role (place \texttt{AnnounceCandidateRole}). Three cases are now possible, each adding its specific delay to the overall response time:
\begin{enumerate}
	\item If the cluster majority is up ($\geq \lfloor C/2 \rfloor+1$) and the replicas acknowledge the candidate as a new leader before the expiration of the \emph{candidate timeout}, the candidate is elected as the leader (output gate \texttt{setLeaderUp}). The announcement of the candidate role from the candidate to the cluster majority takes an additional round trip.
	\item If another leader is identified while the replica is in the candidate state, the candidate replica becomes a follower and a token moves from the \texttt{AnnounceCandidateRole} place to the \texttt{setNewFollowerUp} output gate.
	\item If the cluster majority sends no acknowledgment to the candidate nodes during the candidate timeout (occurs whenever a total of $\lf (C/2) \rf +1 $ replicas are still down), the candidate waits for the timeout to expire and then repeats the candidate procedure.
\end{enumerate}

\begin{figure}[htb]
\vspace*{-0.4cm}
		\centering
		\includegraphics[width=0.45\textwidth]{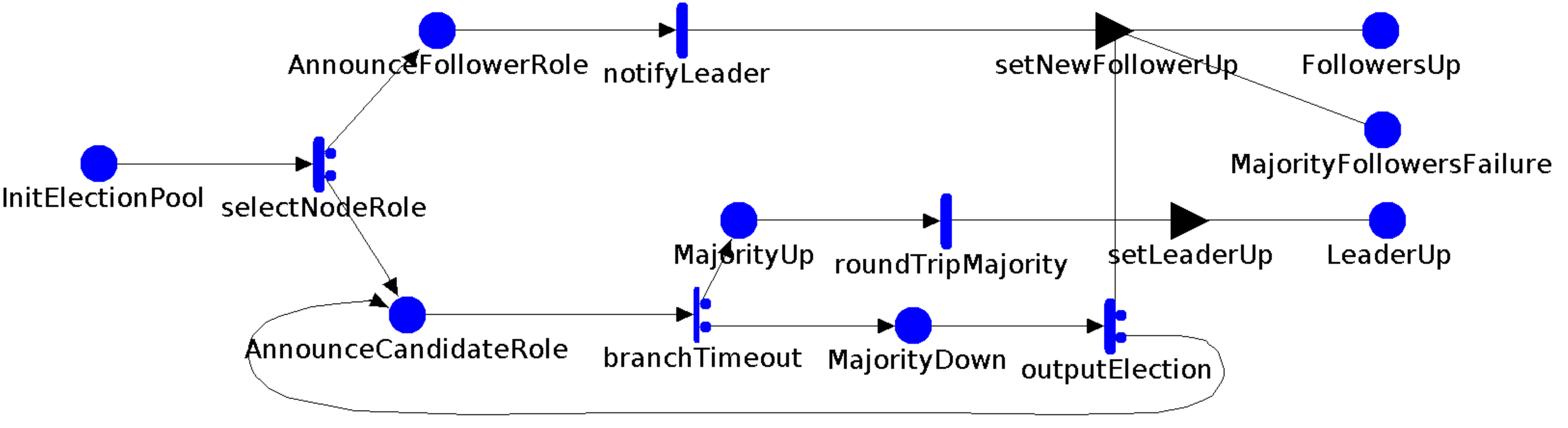}
		\caption{The RAFT recovery SAN model depicts the inclusion of a previously unavailable controller replica into the cluster. Depending on the current state, the replica may become either the new RAFT leader or a follower. Duration of the recovery process will affect the resulting event response time and the cluster availability if the recovering replica is needed to establish a follower-majority and elect a new RAFT leader.}
\vspace*{-0.2cm}
		\label{fig:raftrecovery}
\end{figure}

If the replica becomes a RAFT leader, a token is assigned to the \texttt{LeaderUp} place (previously empty), alternatively the token is assigned to the place \texttt{FollowersUp}. In both cases, the \texttt{NodesUp} token counter is incremented by 1. 


\vspace{-0.3cm}
\section{Evaluation}
\label{evaluationm}
\subsection{Model parametrization using a RAFT experiment}
To evaluate the general fitness of our response time model, we first compare the proposed model against an experimental RAFT setup in a zero-failure scenario. For this purpose, we implement a RAFT agent and deploy multiple copies thereof in a RAFT cluster. For the RAFT backend implementation, we use the open-source Java library \emph{libraft} \footnote{libraft - Raft Distributed Consensus Protocol in Java: \url{https://libraft.io}}. The cluster was organized so that the controller nodes, acting as RAFT agents, were reachable in an any-to-any manner over a single-hop Open vSwitch instance. We configured the Open vSwitch to inject a constant symmetrical delay of $5ms$ on each egress port and we then used this value as a deterministic base leader-follower-majority delay $T_M$ in the model parametrization. Furthermore, based on the \emph{raftlib} performance observations, we modeled the commit delay parameter $T_A$ as an exponentially distributed delay with a mean of $1ms$. The resulting modeled response time and the comparison with the experimental results for different controller cluster sizes are depicted in Fig. \ref{fig:experimental_comparison}. 
As can be noted, our model predicts the expected performance well. To reflect the stochastic performance guarantees when replica failures are concerned, we resort to using only SAN-based analytical modeling for most accurate approximations. 

\begin{figure}[htb]
\centering
\includegraphics[width=0.5\textwidth]{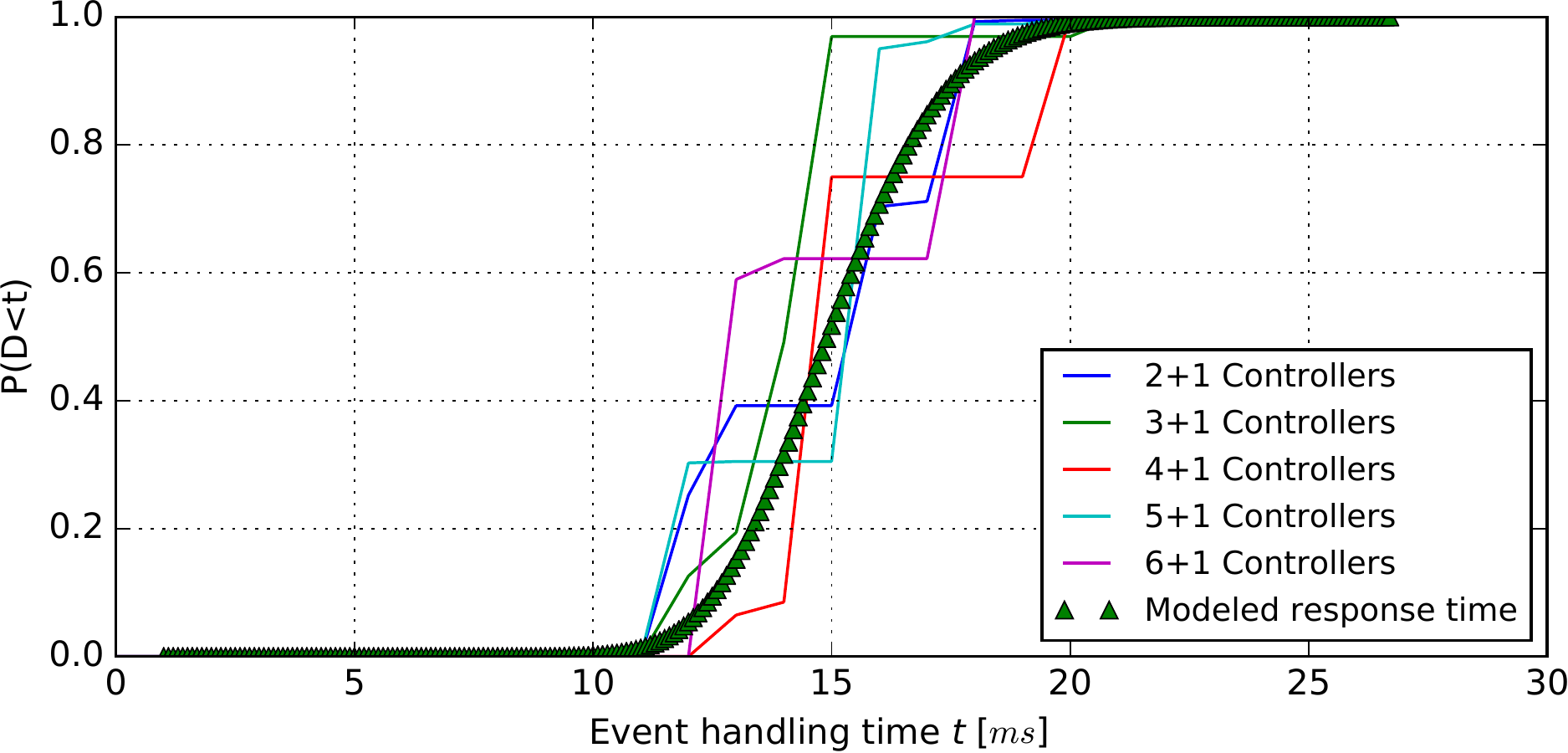}
	\caption{Comparison of experimentally observed and modeled RAFT performance with clusters of various sizes. Represented are the CDFs of per-cluster-configuration measurements, with each measurement encompassing 1000 sequential write operations. The observed delay considers a fixed single-hop packet latency of $5ms$ in between the RAFT leader and replicas, as well as a $1ms$ data-store commit time in the RAFT leader and replica majority. Client and application delays were not considered in this experiment. The measurements were taken in a zero-failure state of the RAFT cluster and should serve as an initial indicator of the response time model fitness.} 
\vspace*{-0.3cm}
\label{fig:experimental_comparison}
\end{figure}

\vspace{-0.3cm}
\subsection{Fast recovery mechanism for bundles and processes}
\label{ssec:recov}
In contrast to evaluating the system with purely fixed software repair rates, as was done in relevant past studies \cite{nencioni2016availability, liu2005proactive}, we utilize a recovery model that reflects much closer the actual state-of-the-art SDN controller implementations. We further propose an optimization to enhance upon the standard repair time in the face of controller failures. We assume a \emph{watchdog}-like mechanism implementation that monitors the critical controller components' health and correctness. The watchdog can monitor both the granular SDN controller applications (bundles) and the actual controller process (that comprises many bundles). Whenever a bundle or a process fails, we assume an immediate scheduling of a rejuvenation procedure that repairs the affected software component. 

\emph{Realization:} While there may exist various designs to realize a watchdog functionality for the purpose of monitoring the liveness of a software or hardware component, we opted to implement the watchdog as a software-agent external to the OSGi container hosting the SDN controller bundles. We deployed the watchdog agent on the same host machine as the monitored SDN controller instance. Following a successful start-up of both the watchdog and the SDN controller processes, the watchdog establishes a connection to the OSGi environment hosting different controller bundles. We make use of the Apache Karaf's\footnote{Apache Karaf - an OSGi distribution offered by the Apache Software Foundation based on Apache Felix - \url{https://karaf.apache.org}} \emph{Remoting} mechanism to allow for remote connections to a running Karaf instance.

Our agent periodically polls the status of a bundle's lifecycle and discovers that the bundle is in one of the following UP states: \{\emph{INSTALLED, STARTING, ACTIVE}\}; or DOWN states: \{\emph{UNINSTALLED, STOPPING, RESOLVED}\}. Upon discovery of a bundle that is in a DOWN-state, the agent schedules a \emph{bundle:start}-transition for the affected bundle, in order to get it up and running in an UP-state. In the case of an unsuccessful remote connection to Karaf, the watchdog evaluates the current list of processes for false positives and, if a missing Karaf process is detected, it schedules an immediate restart of Karaf.

The watchdog process could also be executed externally to the machine running the SDN controller. Hence, while not considered in our evaluation, the same mechanism can be applied to schedule physical or VM reboots in case of a hardware or hypervisor failure. On the other hand, hardware or hypervisor issues may be a sign of misconfigurations or recurring defects whose source should be diagnosed manually.

To collect the accurate real-world repair rates for controller bundles and processes, we have used our watchdog agent implementation to evaluate the bundle and process reboot times in a clustered OpenDaylight (ODL) setup. We had experimentally injected bundle- and process-critical failures in sequence and then measured the subsequent recovery time required to re-stabilize the system. The distinguished mean bundle and software process repair times, measured during the controlled rejuvenation of the critical RAFT component \emph{sal-distributed-datastore} and the ODL's controller process, ticked at $182.9ms$ and $26.9s$ respectively, far below the 3 minute recovery intervals previously proposed in literature \mbox{\cite{nencioni2016availability, verbrugge2005general}}. The measured recovery time purposely does not include the time needed to re-include the recovered node in the RAFT cluster, since this is modeled as a separate non-deterministic process in our SANs. The bundle and process reboots took place inside a dedicated ODL VM that was part of a bigger ODL controller cluster, virtualized on a modern Intel Xeon-based server, with each of the ODL VMs assigned 4 vCores and 8 GB of DDR4 memory. ODL was loading the OSGi bundles available in the \emph{OpenFlowPlugin} and \emph{Controller} projects and had the \emph{Clustering} component enabled.

\begin{table}[htb]
	\centering	
	\caption{SAN model parameters used in our solutions.}
	\begin{tabular}[htb]{ l l l l }
		  \hline 
		  Parameter & Intensity & Unit & Meaning  \\
		  \hline
		  $T_A$ & $1$ & [ms] & Application handling time\\
		  $T_C$ & $1$ & [ms] & Data-store commit delay \\
		  $1/{\lambda_{f}}$ & $225$ & [ms] & Mean follower timeout \\
		  $1/{\lambda_{ca}}$ & $225$ & [ms] & Mean candidate timeout \\
		  $T_{CL}$ & $50$  & [ms] & Client timeout \\
		  $T_R  (T_{M_{worst}})$ & $10$ & [ms] & Worst-case replica-leader delay\\
		  $T_{M_{best}}$ & $5$ & [ms] & Best-case majority-leader delay\\
		  $T_{CR}$ & $1$ & [ms] & Delay client-to-replica\\
		  $N_{F}$ & $1 .. C$ & N/A & Controller failure count\\
		  $1/{\lambda_{F_H}}$ & $6$ & [months] & Hardware failure rate \\
		  $1/{\lambda_{F_S}}$ & $1$ & [week] & Software failure rate \\
		  $1/{\lambda_{F_{S_i}}}$ & $30/\#_{F}$ & [ms] & Software failure rate (injected)\\
		  $1/{\lambda_{R_H}}$ & 12 & [h] & Hardware repair rate \\
		  $1/{\lambda_{d}}$ & $\infty$ & [h] & Critical data plane failure rate \\
		  $1/{\lambda_{R_{S}}}$ & $3$ & [minutes] & Bundle and process repair rate \\
		  $1/{\lambda_{R_{Sbw}}}$ & $182.9$ & [ms] & (\emph{Watchdog}) Bundle repair rate \\
		  $1/{\lambda_{R_{Spw}}}$ & $26.9$ & [s] & (\emph{Watchdog}) Process repair rate \\
		  $E_s$ & $20$ & N/A & Erlang approximation stages \\
		  $R_{M}$ & $10$ & N/A & Max. inconsistent RAFT terms\\
		  \hline
	\end{tabular}
	\label{tab:generalmodel}
\end{table}

\vspace{-0.3cm}
\subsection{On parameter selection}
\label{ssec:param}
To evaluate the expected response time metrics of various cluster configurations, we vary the SDN cluster size and hence the number of controller replicas that take part in the RAFT algorithm as per Table \ref{tab:generalmodel}. The generalized long-term software and hardware failure rates, as well as the hardware repair rates are taken from Liu et al.\cite{liu2005proactive}. As discussed in Subsection \ref{ssec:clusfail}, to allow for granular worst-case response time analysis, we model single and correlated failure injections with varying number of failures, where following a failure, a replica is temporarily excluded from the cluster until recovered. To depict the benefits of failure source differentiation and the proposed watchdog mechanism, we distinguish mixed and software-only failures, and vary the number of failure injections between 1 and $\lf C/2 \rf +1$ (majority nodes down) controller failures. The fact that the process of uniformization may only be applied to exponentially distributed transition rates makes our estimations slightly pessimistic. Thus at the cost of the generated Continuous Time Markov Chain (CTMC) state size and required solving time, we approximate every deterministic message delay and timeout and minimize the total distribution variance using a 20-stage Erlang distribution. 

\vspace{-0.3cm}
\section{Results of our analytical evaluation}
\label{resultss}

In this section, we present the results of the analytical evaluations for various SDN cluster sizes and arbitrary numbers of injected failures. We emphasize the benefits of a fast recovery mechanism for the experienced worst-case response time of an SDN cluster prone to software failures, and visualize its advantages for the long-term system availability. Finally, we discuss the complexity properties of our approach.

\vspace{-0.3cm}
\subsection{Response Time Analysis}
\label{ssec:resultsresp}

When a single random-role controller from the SDN cluster fails as a result of a hardware, process or bundle failure (each being equally probable), deploying a larger number of SDN controller replicas ensures an overall lower expected response time (see Fig. \ref{fig:figure2}). This is related to the probability of a leader being injected with a failure, hence necessitating a leader re-election to move forward the state. The probability of a leader failure becomes increasingly lower when larger clusters are deployed (as explained in Subsection \ref{ssec:clusfail}). 

\begin{figure}[htb]
	\centering
	\includegraphics[width=0.45\textwidth]{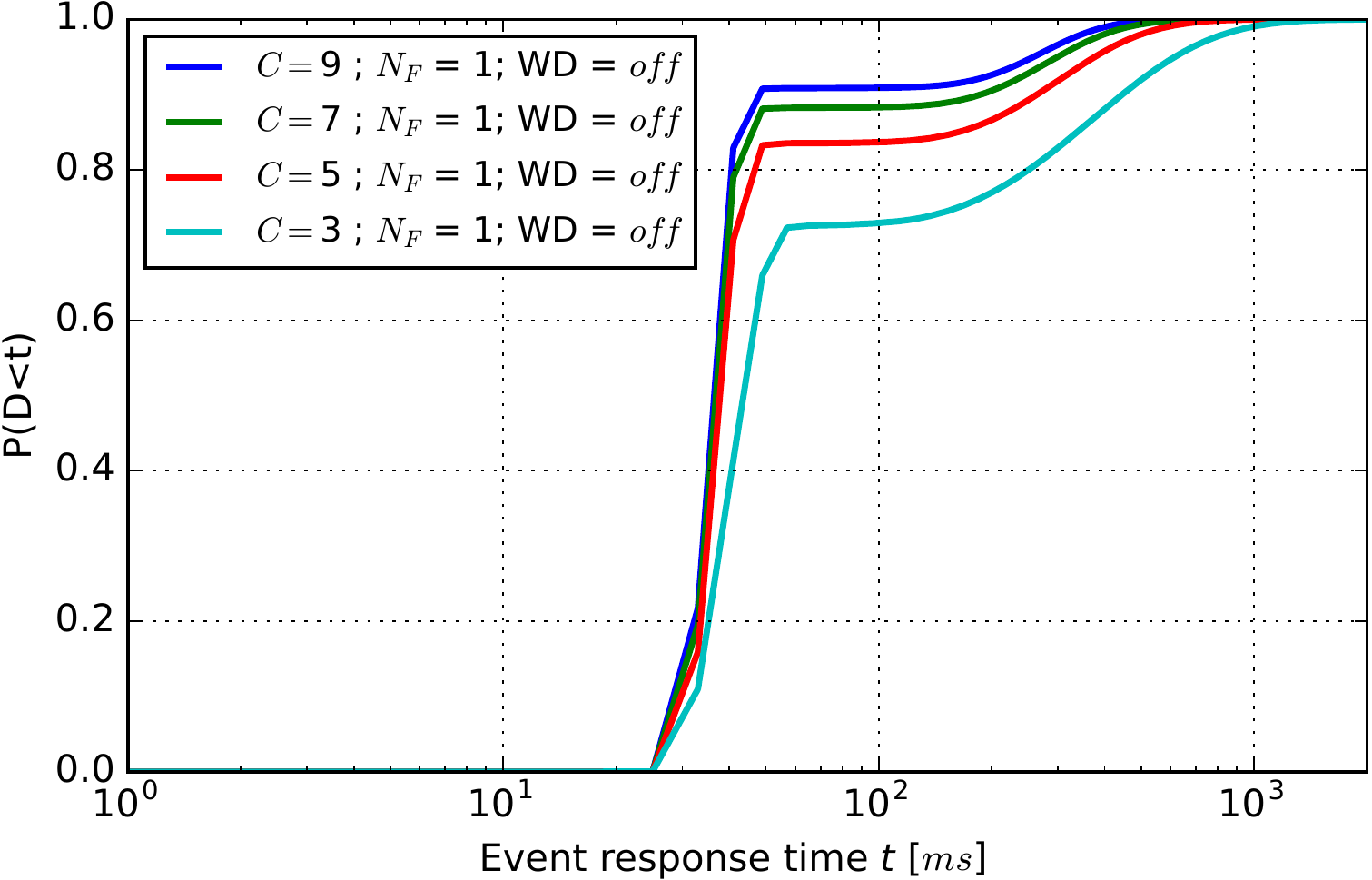}
	\caption{Varying probability of an event being successfully handled in a given time period $t$ for different SDN controller cluster sizes $C$. The probability of the RAFT leader failing is inversely proportional to the cluster size.}
\vspace*{-0.3cm}
	\label{fig:figure2}
\end{figure}

Next, we evaluate the probability of meeting an event handling deadline when the majority of nodes in the cluster have failed. The expected response times where mixed hardware and software failures, as well as exclusively bundle-level failures may occur, are depicted with and without the watchdog (\emph{WD}) mechanism enabled in Fig. \ref{fig:f1} and \ref{fig:f2}, respectively. The watchdog mechanism enables faster recovery of replicas and hence faster repeated processing of an event in the case of leader and follower majority failures. An SDN cluster equipped with the watchdog mechanism on average processes the events faster and with a higher probability than the one without. Especially when simultaneous hardware failures are improbable and software failures are typical, the fast software recovery provides obvious response-time benefits (Fig. \ref{fig:f2}). 

\begin{figure}[htb]
	\centering
	\subfloat[Resulting response time assuming an occurrence of $N_F$ combined correlated hardware and software (process, bundle) failures. All three types of failures are injected with equal probability. ]{\includegraphics[width=0.5\textwidth]{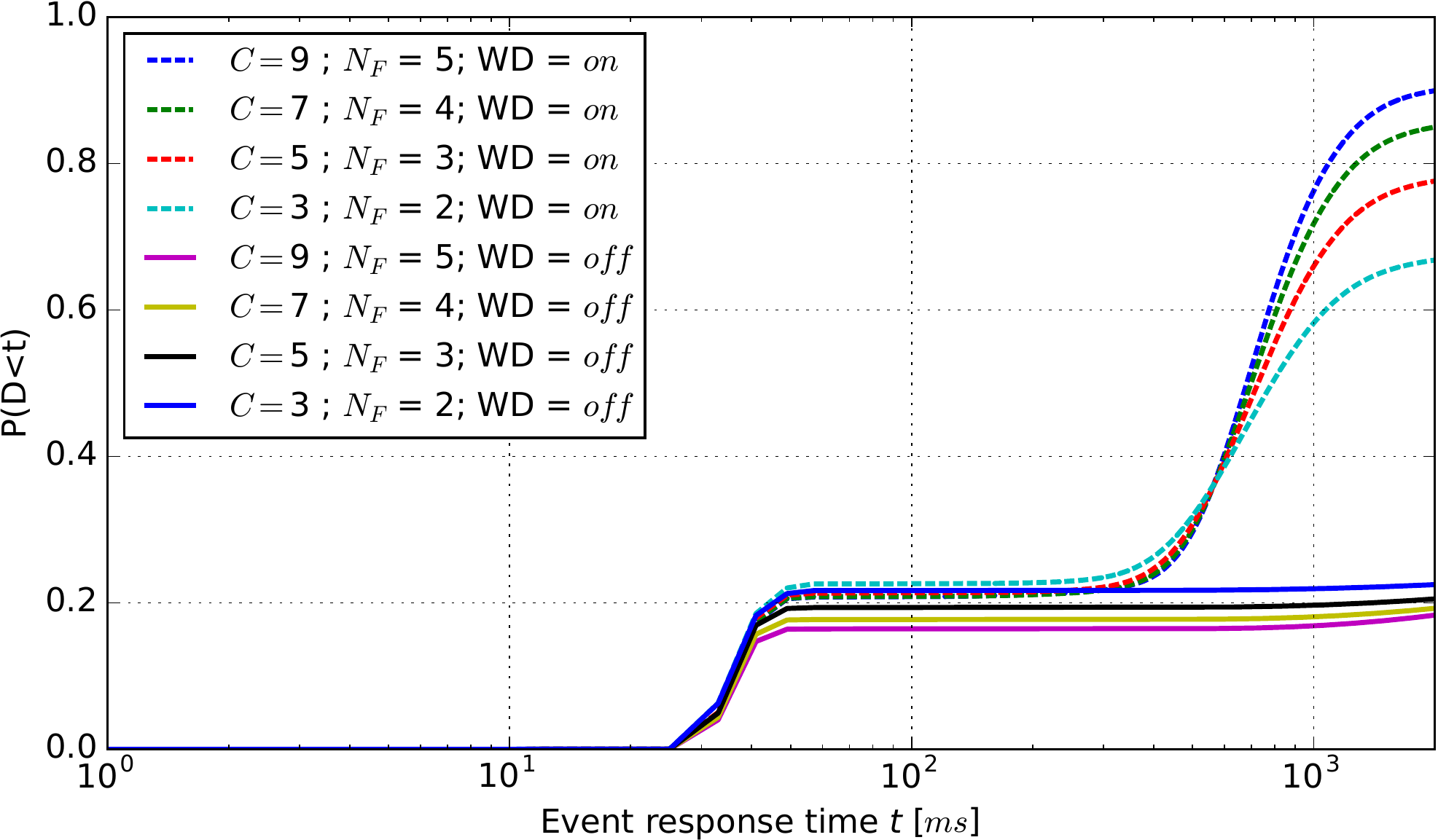}\label{fig:f1}}
	\hfill
	\subfloat[Resulting response time assuming $N_F$ bundle-only failures. The watchdog mechanism will guarantee a timely repair and inclusion of the recovered RAFT node in the cluster.]{\includegraphics[width=0.5\textwidth]{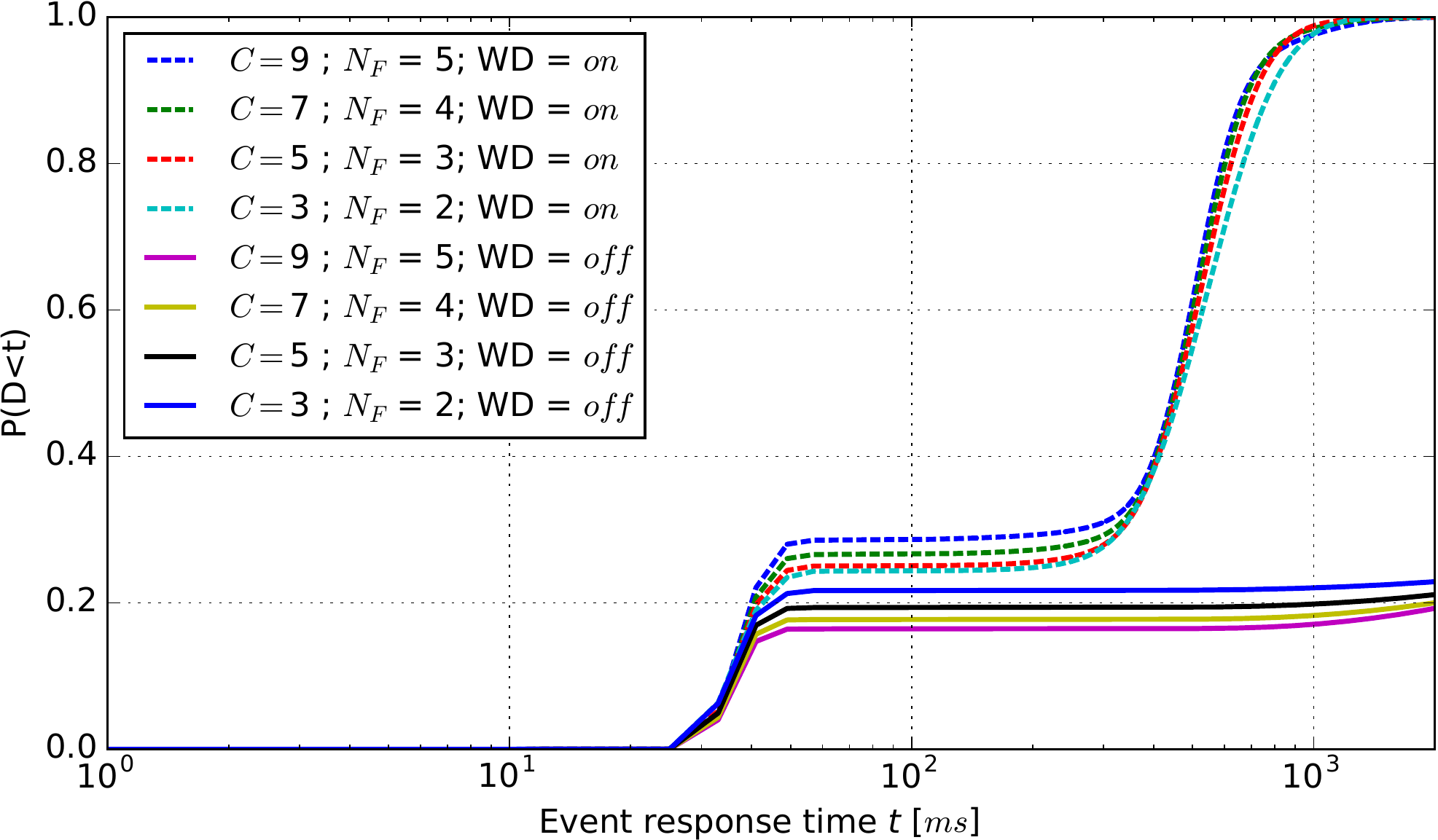}\label{fig:f2}}
	\caption{Probability of receiving an event response during an observation window, assuming a simultaneous occurrence of (a) $N_F$ mixed and (b) $N_F$ software-bundle only controller failures in a cluster comprised of $C$ controllers. The failures are injected at rate $N_F \cdot 0.0333$ (all correlated $N_F$ failures are thus expected to be injected by time point $t=30ms$).}
\vspace*{-0.3cm}
	\label{fig:figure3}
\end{figure}

Fig. \ref{fig:arbitraryfailures} depicts the effect of the consecutive failures on the experienced response time in a 7-node controller cluster. If the majority nodes remain available after each individual failure, the time to respond is governed by the case where a cluster leader fails and a new leader election procedure is automatically initiated. There is no noticeable difference in the convergence time regardless of the (non-)usage of the watchdog mechanism in this particular case. The lower the maximum number of induced failures induced, slightly shorter is the expected response time. This may be related to the fact that the follower timeouts are exponentially distributed, hence a higher number of active nodes that time out after a leader failure leads to an overall lower expected time to select a candidate and repair the cluster.

\begin{figure}[htb]
	\includegraphics[width=0.5\textwidth]{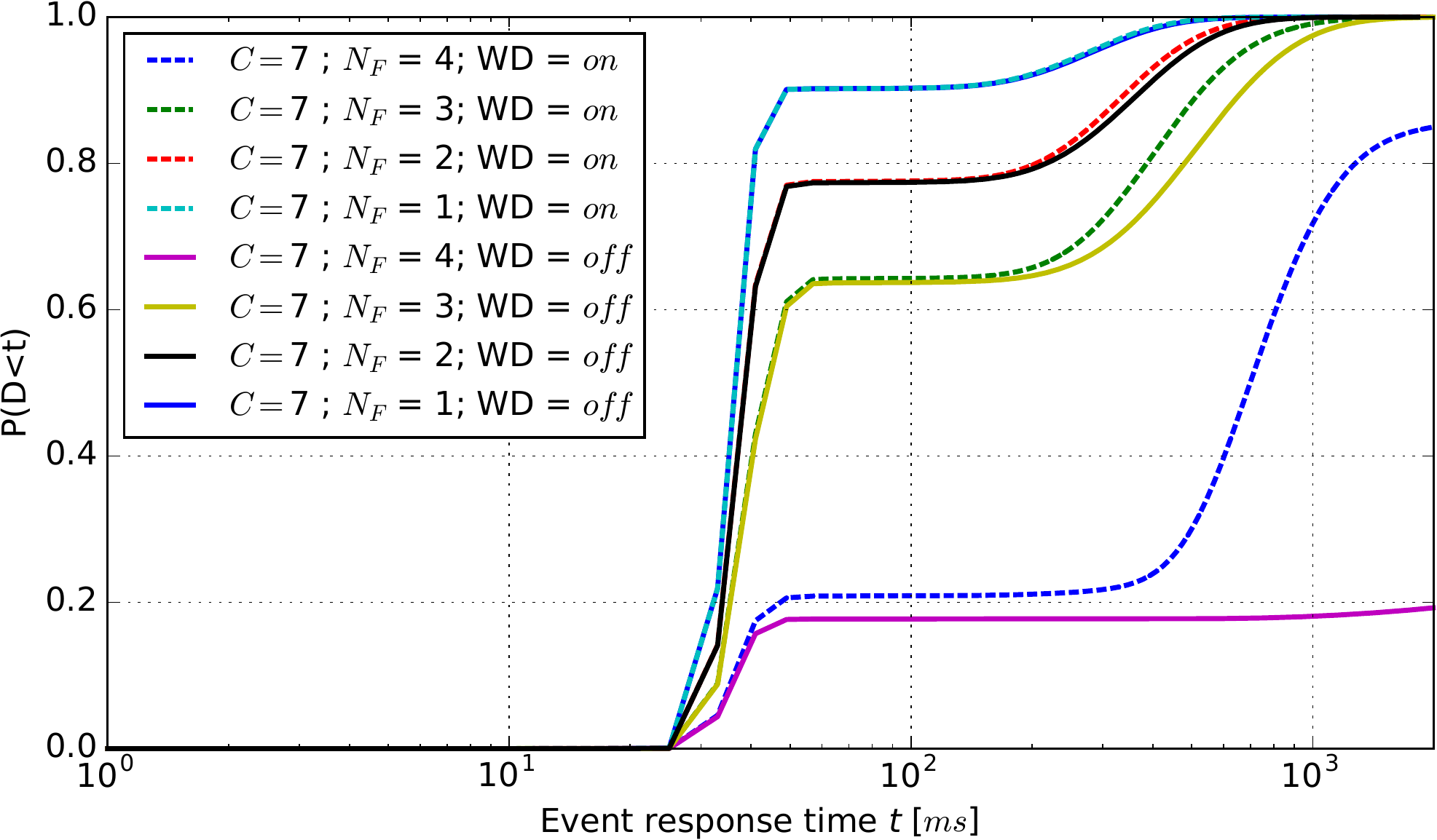}
	\caption{Resulting response time assuming an occurrence of $1 \leq N_F \leq (\lfloor C/2 \rfloor +1)$ controller failures in a 7-node controller cluster. The response time is governed by the duration of the leader election procedure. When the majority of controllers are unavailable, the usage of the \emph{watchdog} mechanism (dashed) leads to important benefits w.r.t. expected worst-case response time.}
\vspace*{-0.3cm}
	\label{fig:arbitraryfailures}
\end{figure}

\vspace{-0.3cm}
\subsection{Cluster Availability}
\label{ssec:availabilityresult}

Next, we emphasize the long-term advantage of an SDN controller bundle/process watchdog mechanism by evaluating the availability of a 3-node cluster configuration over an observation period of 1000 hours. Fig. \ref{fig:figure4} depicts the unavailability of a 3-node controller cluster setup. We define the unavailability measure as the probability of encountering an unavailable cluster of controllers at any time instant $t$ as $P_{C_U}(t) = 1-P_{C_A}(t)$. Here $P_{C_A}(t)$ represents the probability of encountering a system in a state where the RAFT leader and the majority of RAFT followers are available and have converged their leader-election processes. Software and hardware failures are modeled using the long-term exponential hardware and software failure rates presented in Table \ref{tab:generalmodel}. 
The approximated unavailability measure saturates after $\sim$85 hours, which is an expected mean failure time for the combined software failures at bundle and process level, given the individual exponentially distributed failures with a mean of $1$ week ($\sim$170 hours) for individual arrivals. We consider the process and bundle failure arrivals as two independent Poisson processes with variably configured rates. Hence, merging the two independent processes with equal arrival rates results in an approximately halved inter-arrival time between software failures. The usage of a watchdog that proactively rejuvenates a system after a software failure leads to a shorter overall experienced downtime, and hence a lower expected RAFT cluster unavailability in the long-term. Configurations with five or more replicas guarantee a negligible unavailability of $<1e^{-9}$ and are hence not included in the figure. 

\begin{figure}[htb]
	\centering
	\includegraphics[width=0.5\textwidth]{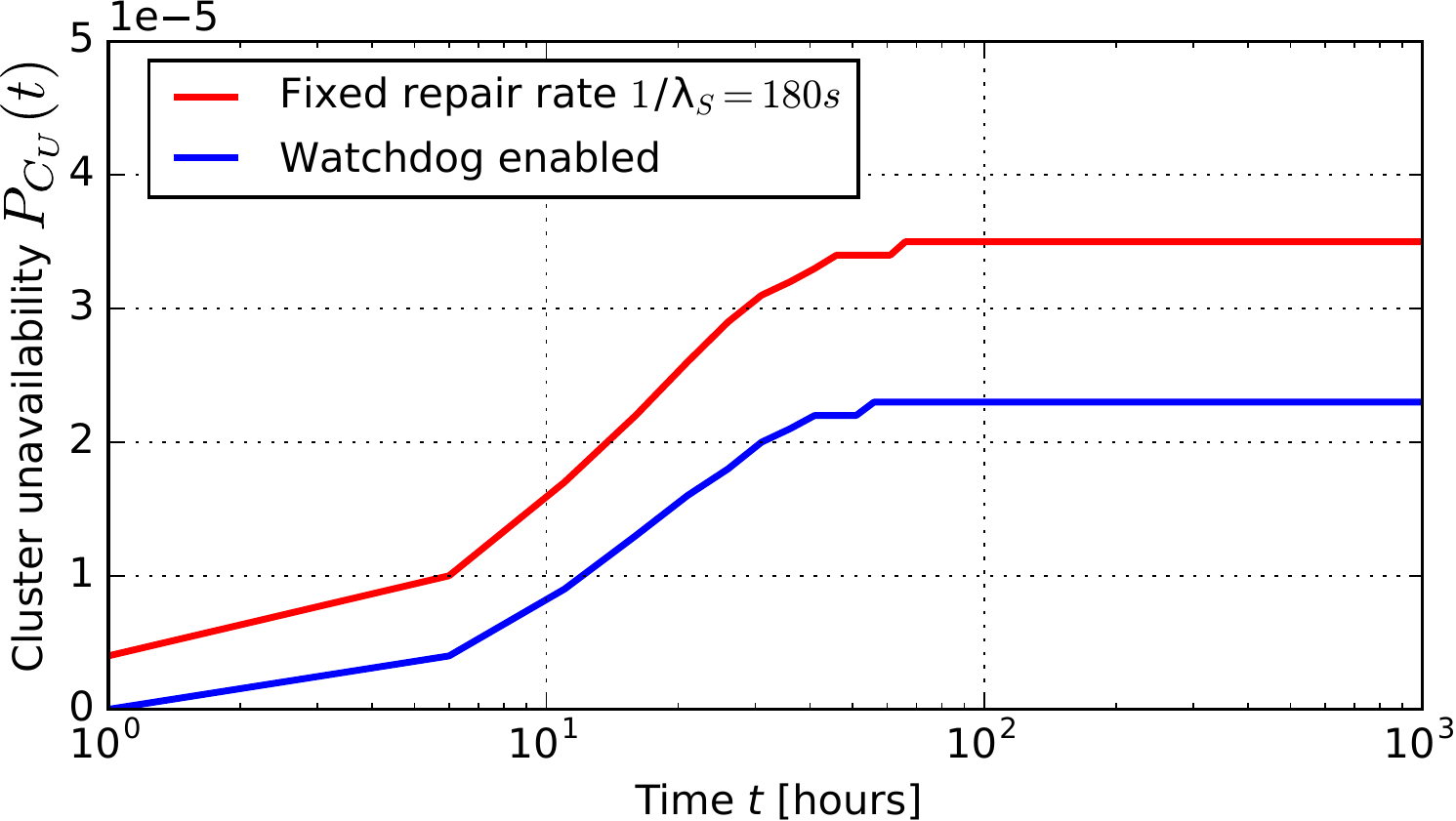}
	\caption{Transient analysis of the SDN controller cluster unavailability over a period of 1000 hours. The cluster size of exactly three controllers was considered in the transient analysis. As expected, the inclusion of a liveness guard mechanism results in a lower overall expected unavailability. SDN controller clusters that include five or a higher number of replicas per cluster have shown to posses negligible availability concerns. This confirms the claims made in \cite{nencioni2016availability}, where authors discuss the minimal effect of long-term failure rates on the experienced downtime of an SDN control plane.}
\vspace*{-0.4cm}
	\label{fig:figure4}
\end{figure}

\vspace{-0.3cm}
\subsection{Model Complexity and Solve Time}
\label{ssec:scalability}
Compared to the manual Markov Chain modeling, SANs allow for more compact modeling of complex scenarios. Analytically, both options need to solve the same CTMC and have to deal with an exponential increase in model size which may result in inefficient or intractable analytical solutions when complex models are concerned \cite{sanders1991reduced, heegaard2009survivability}. The model complexity dictates both the amount of computational resources and the time required to solve the model.

Fig. \ref{fig:figure5} depicts the state space sizes of the generated CTMCs. The generated state space is used by the transient solver to find the transient solutions for short-term ($N_F$ lower than $C$) and long term ($N_F$ considers up to $C$ failures) numerical analysis. The model complexity increases with the number of possible combinations the system may occupy. For short-term response time analysis we limit the complexity of the model by considering only the injected correlated failures - this is realistic as only a very short time period ($1s<x<2s$) is considered (see Figures \ref{fig:figure2} and \ref{fig:figure3}). For long-term analysis, additional system states, where more than just the majority of nodes may fail could be of interest (consider Fig. \ref{fig:figure4}). Fig. {\ref{fig:figure5}} shows the CTMC state space sizes for the cluster configurations up to $C=19$. We observe that, for some parametrizations, the compiled state space size grows exponentially with the number of controller replicas. The number of possible failure injections dictates the number of generated unique combinations. For the most accurate setting of the $E_S = 20$ (20 Erlang stages, see Subsection {\ref{ssec:param})} and cluster sizes of 17 and more replicas, we have encountered memory handling limitations in the flat state space generator in M\"obius. Namely, if the solution should cover for all theoretically possible system combinations, i.e. when failure of every single node should be considered, the solution space eventually grows to an intractable amount of states for very large cluster sizes. To cater for the scalability of our solution when analyzing large control planes, we propose three options:
\begin{enumerate}
	\item State space largeness avoidance by applying a scenario-based approach to the worst-case modeling. For example, one could consider a limited number of maximum failure injections. By limiting the number of maximum failure injections to $N_F = \lfloor C/2 \rfloor +1$, large-scale clusters can be analyzed successfully (see Fig. {\ref{fig:figure5}}).
	\item State space largeness avoidance by trading solution accuracy, e.g. by manipulation of the Erlang stages used for the approximation of the deterministic transitions.
	\item Faster convergence of the transient solver by raising the error tolerance of the uniformization (see Equation 1).
\end{enumerate}

\begin{figure}[htb]
	\centering
	\includegraphics[width=0.5\textwidth]{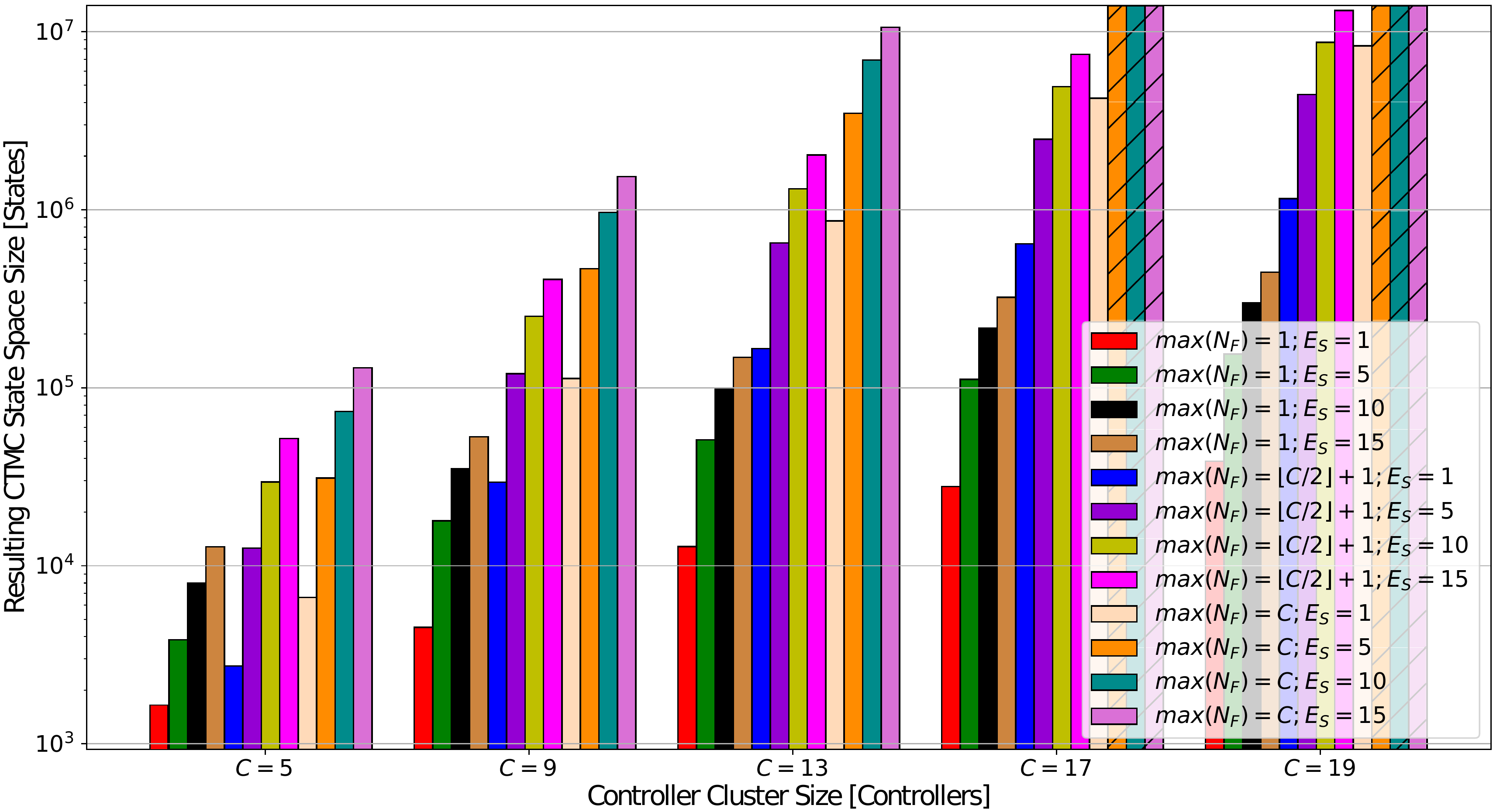}
	\caption{Size of the CTMC state space generated using the SAN models and parameters discussed in Sections III and IV, respectively. The lower the number of controller failures of interest (i.e. where $N_F<C$), the smaller the resulting CTMC state space size. If the possibility of an eventual occurrence of failures in all nodes is assumed, the state space grows correspondingly, reaching up to $10^7$ possible state space combinations with controller cluster size set to $C=13$ and the maximum accuracy $E_S=15$. Striped bars represent the unsuccessful CTMC compilations where the flat state space generator fails to compile the state space. However, by considering a lower number of Erlang approximation stages $E_S$, $C=19$ and more controller replicas can be handled with a limited inaccuracy (see Fig. {\ref{fig:erlangstages}}). Similarly, a focused assumption on the maximum number of possible failure occurrences helps the scalability of the solution (where $max(N_F) < C$).}
	\label{fig:figure5}
\end{figure}

For completeness, we also evaluate the second option by varying the Erlang stage parametrization. We take note of the effect on the overall result accuracy for the transient analysis of a 7-node cluster. Fig. {\ref{fig:erlangstages}} depicts the inaccuracy of the latency bound introduced by lowering the number of Erlang stages from $E_{S_{high}} = 20$ to $E_{S_{low}} \in \{5, 10, 15\}$. At $E_S = 5$, the generated state space size is decreased by a magnitude (see Fig. {\ref{fig:figure5}}) and is hence, in addition to the first option, an effective method of deploying our models in a scalable manner. From this study, we conclude that the state space generation process is scalable as long as the accuracy and failure injection parameters are selected carefully for the use case at hand.

\begin{figure}[htb]
	\centering
	\includegraphics[width=0.45\textwidth]{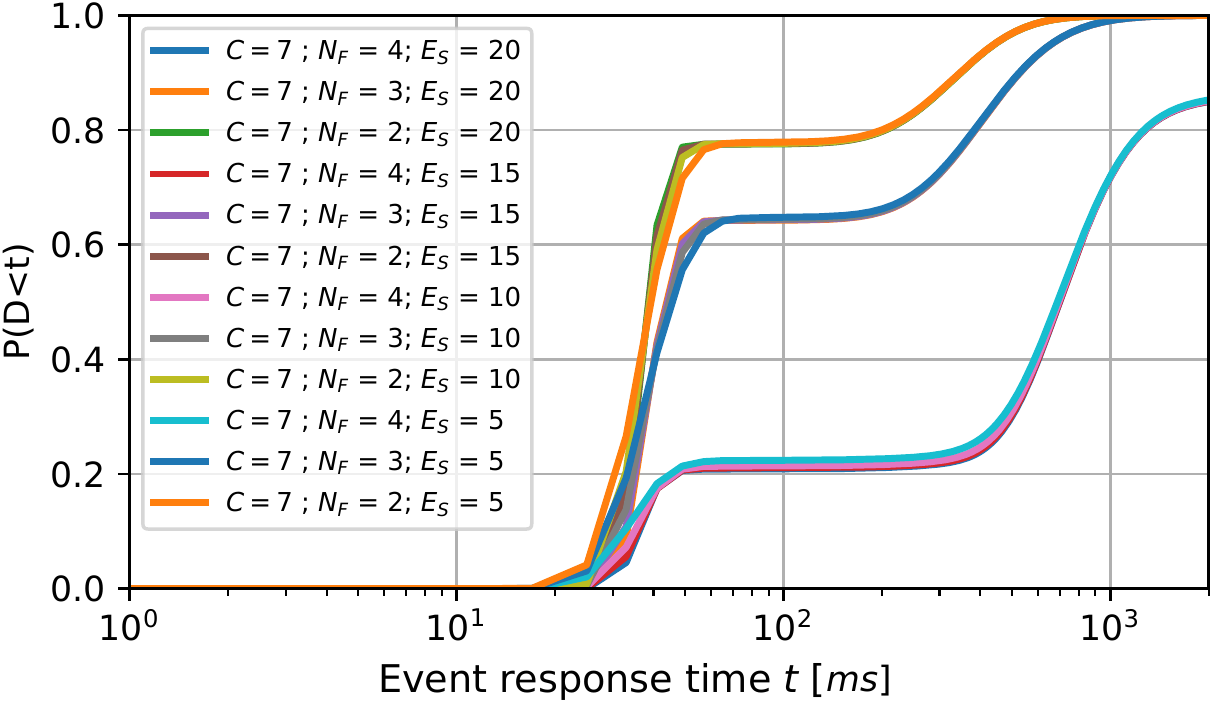}
	\caption{Inaccuracies stemming from a decreased number of Erlang stages $E_S$ used in the approximation of deterministic transitions are negligible. Inaccurate approximation of a deterministic distribution lead to a higher variance for the random variable describing the failure arrivals. Hence, for small $E_S$ the solver estimates a more relaxed (thus more pessimistic) latency bound.}
\vspace*{-0.2cm}
	\label{fig:erlangstages}
\end{figure}

We next consider the performance overhead of the state space generation in our approach. Fig. {\ref{fig:overhead}} shows how the scenario where $max(N_F) = C$ with $C=19$ and $E_S = 5$ results in a tolerable $\sim10^3$ seconds solving period.

\begin{figure}[htb]
	\centering
	\includegraphics[width=0.45\textwidth]{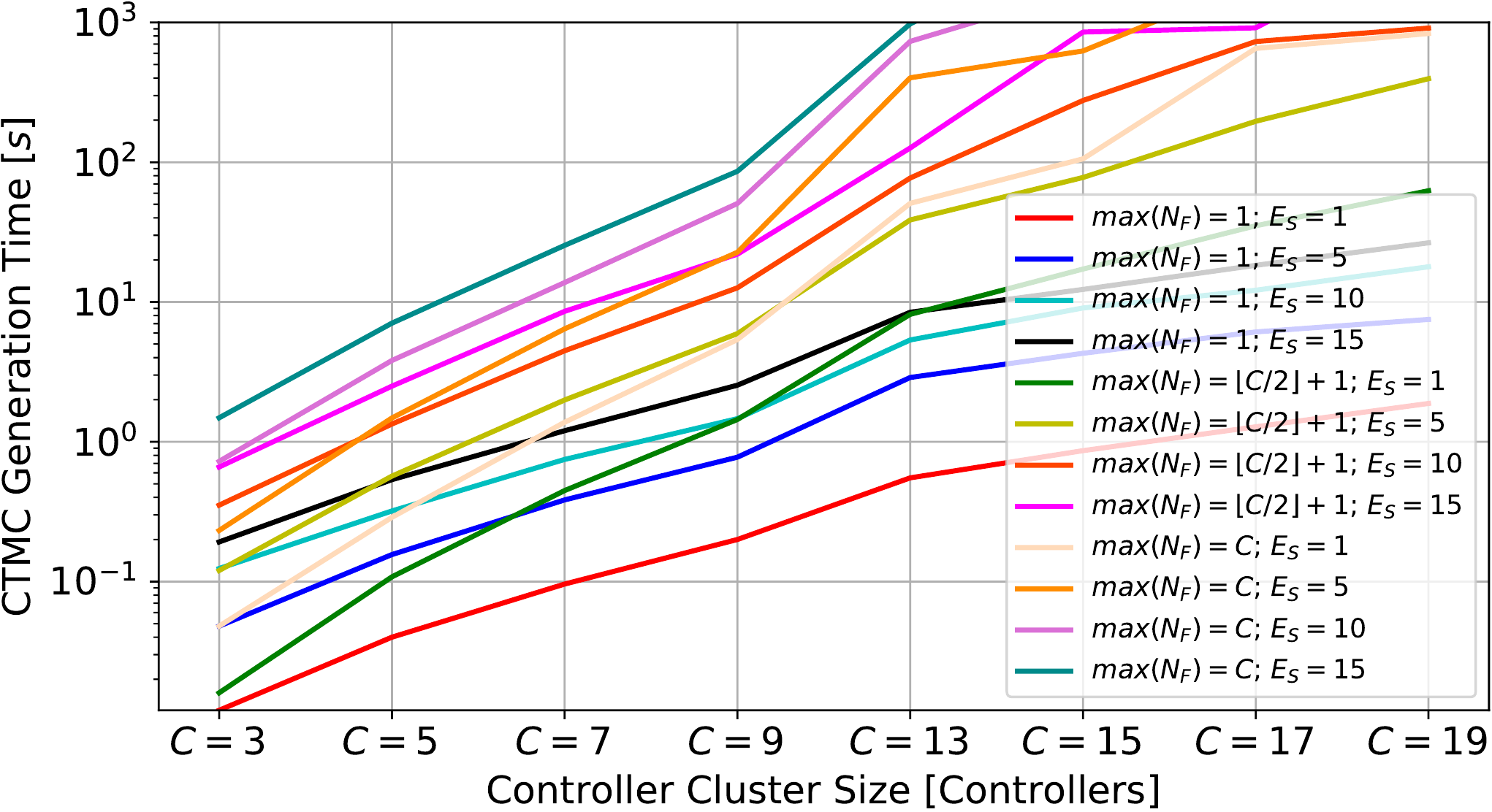}
	\caption{Overhead of the CTMC compilation for varying cluster sizes $C$, failure injection counts $max(N_F)$ and Erlang parameterizations $E_S$. While very accurate and large-scale combinations may lead to intractable solutions, feasible solutions can be presented even for the complex deployments of $C=19$ controllers with various degrees of accuracy and all $max(N_F)$ combinations.}
\vspace*{-0.4cm}
	\label{fig:overhead}
\end{figure}

Fig. \ref{fig:figure6} depicts the computation time to solve the presented SANs. The duration of the solution computation of SAN will vary depending on the model complexity (state space size), the definition of the observed performance variable (reward function and the number and granularity of time measurements), as well as the required accuracy and model \emph{stiffness} (the range of the expected action completion times)\cite{malhotra1994stiffness}. In the M\"obius modeling tool, the accuracy of the transient solver indicates the degree of accuracy that the user wishes in terms of the number of decimal places. The solver execution times depicted in Fig. \ref{fig:figure6} were observed for the accuracy parameter set to 9 and an observation window of 1 second (1000 data points). The largest generated state space for the purpose of modeling the largest cluster size necessarily leads to the longest solution computation times. For the analysis scenarios described here, these computation times are feasible.

\begin{figure}[htb]
	\centering
	\includegraphics[width=0.45\textwidth]{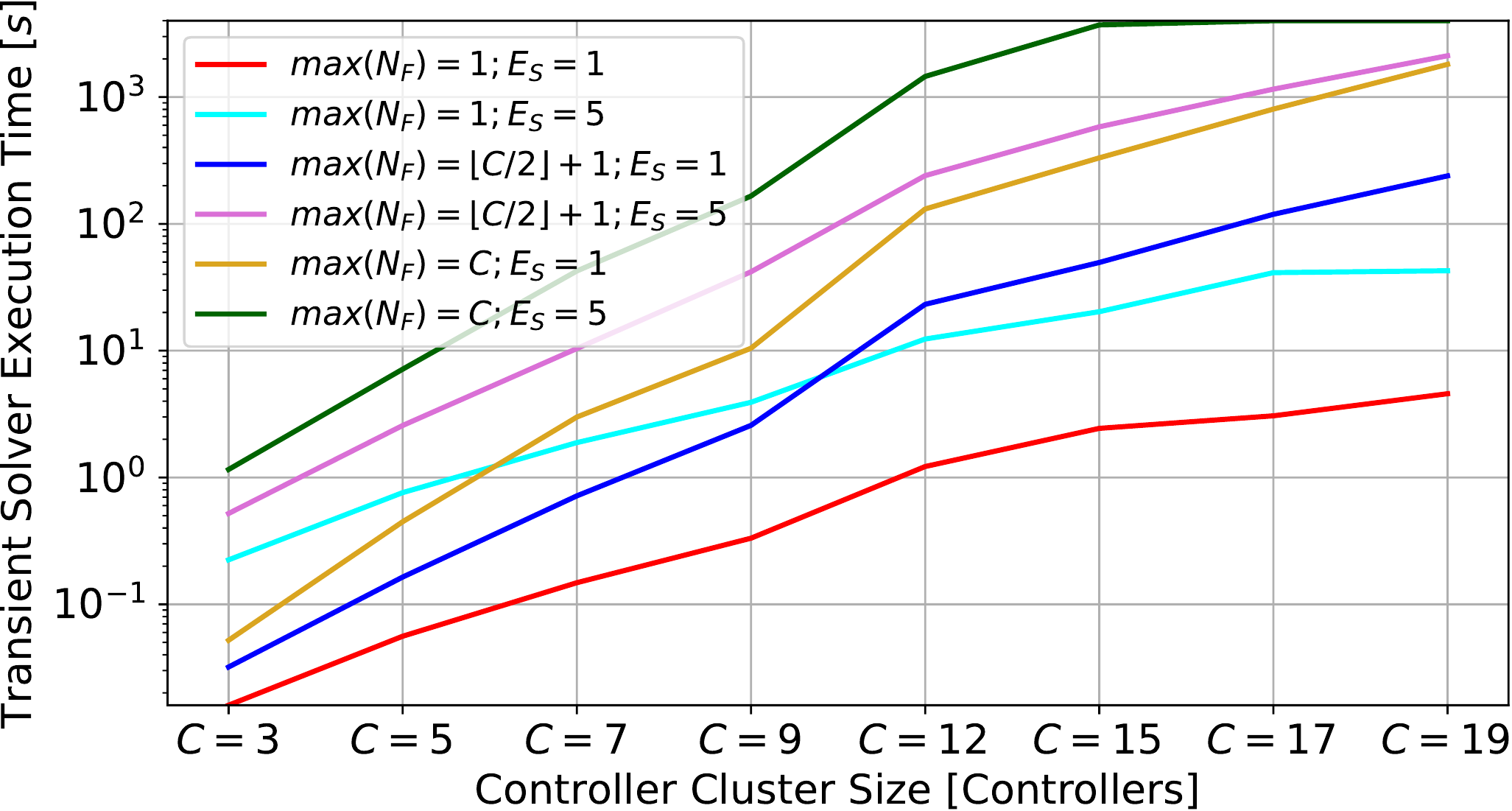}
	\caption{Computation time of the \emph{instant of time} \mbox{\cite{mobius}} transient solutions for the state space sizes depicted in Fig. {\ref{fig:figure5}}. The solution covers the target observation interval of $1$ second at millisecond resolution - hence the transient solver has computed the solutions for 1000 time-points. The computations were executed on a commodity hardware equipped with a modern AMD processor and 32GB of DDR4 memory. The required computation overhead for the numerical solution is feasible for a short-term response time study.}
\vspace*{-0.4cm}
	\label{fig:figure6}
\end{figure}

\vspace{-0.3cm}
\section{Related work}
\label{relatedwork}
In state of the art literature, availability and overhead modeling of SDN has recently started to gain traction. In \mbox{\cite{nencioni2016availability}}, the authors investigate the impact of operational and management failures on the availability in SDNs. They focus on the long-term availability impact of adding additional controllers, but do not include any response time analysis nor consider the impact of controller synchronization at micro-scale.

Tuncer et al. \mbox{\cite{tuncer2015placement}} enhance a controller placement heuristic to cater for the optimality of the controller-network device cluster unbalance. Given an arbitrary network topology, their objective is to compute the number of controllers and the fitting placement, as well as to declare the controller-device assignments when considering a \emph{distance} (e.g. delay) constraint. While the controller-switch assignment was specifically targeted in their work, the same solution could be applied to planning an efficient controller cluster configuration. The problem we solve is complementary to this, since we allow for analyzing \emph{any} given SDN cluster with regards to its worst-case control plane performance at runtime.

Muqaddas et al. \mbox{\cite{muqaddas2016inter, muqaddas2017inter}} investigate the load overhead of the intra-cluster communication in a 2- and 3-controller ONOS cluster. They propose a model to quantify the traffic exchanged among the controllers and express it as a function of the network topology. They did not consider the effect of the transient failures on the response time and availability.

In \mbox{\cite{zhang2016role}}, Zhang et al. describe the \emph{single-data ownership} organizational model implemented by the RAFT algorithm and propose an estimation formula for approximating of the flow setup time in a distributed SDN controller cluster. Their estimation is however fairly simplistic as it models only the average case. The worst-case estimations were not considered.

Ongaro \mbox{\cite{ongaro2014consensus}} and Howard et al. \mbox{\cite{Howard:2015:RRW:2723872.2723876}} provide initial performance evaluations of the RAFT consensus algorithm. Howard et al. \mbox{\cite{Howard:2015:RRW:2723872.2723876}} further implement an event-driven framework for prototyping of RAFT using experimental topologies. Contrary to the analytical approach presented in our work, their performance evaluation of RAFT is based on a limited number of repeated experiments and focuses on evaluating the RAFT leader re-election procedure following a failure. Unfortunately, these works do not provide a good understanding of how the overall system response time is affected after a failure.

In two experiment-based studies, Suh et al. \mbox{\cite{suh2016performance, suh2017toward}} measure the throughput and the recovery time of a RAFT-enabled SDN controller cluster with 1, 3 and 5 replicas. They put special focus on the effect of $\phi$ accrual failure detector \mbox{\cite{hayashibara2004spl}} on the resulting performance footprint. The authors deduce that the controller failover time increases as the value $\phi$ increases. With higher $\phi$, the OpenDaylight cluster becomes more conservative in determining a controller failure, hence in case of failures, using a large $\phi$ values will generally lead to slow failure discovery. Authors varied $\phi$ and measured the lowest recovery time of $\sim2,6s$, which is a non-satisfying recovery time for many critical industrial applications. Instead of using an adaptive scaling factor $\phi$, we rely on a fixed \emph{follower timeout} variable with a mean of $\sim225ms$. We assume that the controller-to-controller delays are bounded and will hence not exceed this value except in the case where a controller failure has occurred. This value is recommended by the authors of the RAFT consensus algorithm, and was determined to be a good trade-off between the recovery time and the signaling overhead in their experiments \mbox{\cite{ongaro2014search}}. 

The introduced watchdog mechanism for fast software system recovery relates to the concept of software rejuvenation. Several works have investigated the phenomenon of "software aging" wherein the health of a software system degrades with time \mbox{\cite{xie2004software, vaidyanathan2001analysis}}. These papers conclude that a mechanism which "rejuvenates" or "recovers" the software component to its stable state, would provide long-term benefits in terms of experienced system availability. In this work, we evaluate the benefits of the \emph{reactive} controller recovery where, following a detected controller bundle or process failure, the affected component is reinitialized in order to minimize the downtime. 

Machida et al. \mbox{\cite{machida2014job}} analyze the completion time of a job running on a server that is affected by \emph{software aging}, and consider the benefit of the \emph{preemptive-resume} operation, where a job resumes execution from the point of interruption as soon as the failed server recovers. Similar to this work, we investigate the job completion time for a client request, but we consider a distributed multi-server operation. We focus on the strategy where, assuming a failure occurs, the request is handled from the beginning instead of delegating it to the next server. 


Apart from RAFT, Paxos\mbox{\cite{lamport1998part}} is another influential \mbox{\cite{bolosky2011paxos, burrows2006chubby}} consensus algorithm that eventually motivated the development of RAFT. Paxos ensures that any two distributed servers that are part of the same cluster may never disagree about the value of a particular update, for any applied update in the update history. In its optimizations, its performance is comparable to RAFT, in that, assuming a stable cluster leadership, committing a cluster-wide update takes a single round trip in most cases \mbox{\cite{moraru2013there}}. 
Multi-Paxos \mbox{\cite{lamport2001paxos, lamport1998part}} is a prominent variation of Paxos, that assumes a stable leader for an infinite number of sequential cluster updates. This allows for one-round-trip delay as the first phase of Paxos becomes unnecessary for the majority of updates. In \mbox{\cite{dumulti}} the authors evaluate an implementation of Multi-Paxos and conclude that the overall performance of Multi-Paxos is limited by the slowest node in the fastest cluster majority. This is a valid observation for any quorum-based consensus algorithm, hence we distinguish the leader failures as critical for our analysis. 

To minimize the effect of single-leader failure and maximize the load balancing of requests, Mencius \mbox{\cite{mao2008mencius}} proposes a round-robin-based update-handling by multiple leaders in a Paxos cluster. While it enables higher throughput in the stable case, the cluster will always run at the speed of the slowest elected leader as the new updates may be dependant on previous updates that are assigned to be handled by the slow node. EPaxos \mbox{\cite{moraru2013there}} is a recent leader-less take on Paxos that tries to circumvent these issues. It keeps track of the ordering and mutual dependencies between the client-initiated updates. Hence, it is able to parallelize multiple update instances when no collisions between concurrent client updates are expected. Like RAFT, it requires a single round-trip in most cases to commit a state update, and two round-trips if dependency conflicts arise. Contrary to RAFT and other leader-based Paxos variations, the response time in an EPaxos cluster does not suffer from unstable leaders since the clients may always fall-back to any remaining live leader replica. However, the algorithm adds additional complexity in state-keeping and log compaction tasks because of the added dependency trees. 

Since all available SDN cluster implementations focus on a single master for any switch in its administrative domain at runtime, we put focus on the evaluation of a single-leader RAFT-based cluster and consider its direct comparison with multi-leader EPaxos \mbox{\cite{moraru2013there}} and \emph{eventually consistent} approaches \mbox{\cite{Sakic, levin2012logically}} as future work.

\section{Conclusion and Outlook}
\label{conclude}
SDN enables the necessary control plane robustness by controller clustering and state replication. However, this replication incurs additional performance overhead. Indeed, it is not always clear which particular cluster configuration would best suit the application and network configuration at hand. Existing performance studies of distributed SDN control plane neglect the cluster's response time and availability metrics. 

Hence, we hereby propose the usage of Stochastic Activity Networks (SANs) for modeling and numerical evaluation of distributed SDN clusters. We put special focus on the practically relevant distributed consensus algorithm RAFT, but generalize our model to be applicable to similar Paxos variants (e.g., Multi-Paxos). RAFT is implemented in two dominant open-source SDN platforms and is of practical relevance for performance analysis of the distributed SDN control plane. We introduce and discuss the SAN-based models for response time and availability evaluation, and include a failure injection model for evaluation of the two metrics under the effect of an arbitrary number of correlated controller failures. Using transient solver methods, we are able to provide a probabilistic guarantee on the event handling response times and numerically evaluate the availability property of arbitrary SDN cluster configurations. We have shown that, assuming a balanced distribution of controllers in the network w.r.t. the controller-to-controller delays, larger clusters provide lower worst-case response times and higher system availability. With the help of analytical modeling, the evaluation and optimization of cluster configurations, in order to determine the best suited configuration for the network at hand, becomes possible without costly hardware setups for experimental evaluation or lengthy simulation runs. Analytical modeling further provides for corner case inclusion and tighter stochastic guarantees than possible using experimental sampling. 

Finally, we have proposed the \emph{watchdog} mechanism for fast recovery from software failures in a distributed SDN controller setting. Using transient solvers, we have proven its benefits on the short-term response time and the long-term availability properties of a controller cluster. The solutions to our models are computationally feasible for both the typical (3-5 controllers), and very complex clusters ($\sim$20 controllers). 

Extensions to support the novel leaderless Paxos variants, such as EPaxos, require significant changes in the models used and are thus considered as future work. Furthermore, we provide the response time metrics for a model that assumes an accumulated distribution of RAFT state updates in the latency- and throughput-optimized, batched mode. Extending the proposed model to support a sequential distribution of updates at high scale is non-trivial using SAN-based modeling, because of the added state size complexity. Supporting the queueing behavior when handling client-generated events requires inclusion of additional concepts from classic queueing theory or network calculus for practical value.

\vspace{-0.4cm}
\section*{Acknowledgment}

This work has received funding from the EU's Horizon 2020 research and innovation programme under grant agreement number 671648 VirtuWind and in parts under grant agreement number 647158 FlexNets (by the European Research Council).


%
\vspace{-0.3cm}
\bibliographystyle{IEEEtran}
\bibliography{IEEEabrv,qos}

%


\vspace{-1.2cm}
	\begin{IEEEbiography}[{\includegraphics[width=0.9in,height=1.25in,clip,keepaspectratio]{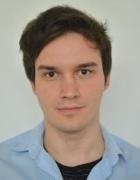}}]{Ermin Sakic} (S$'$17) received his B.Sc. and M.Sc. degrees in electrical engineering and information technology from Technical University of Munich in 2012 and 2014, respectively. He is currently employed at Siemens AG as a Research Scientist in the Corporate Technology research unit. Since 2016, he is pursuing the Ph.D. degree with the Department of Electrical and Computer Engineering at TUM. His research interests include reliable and scalable Software Defined Networks, distributed systems and efficient network and service management. 
\end{IEEEbiography}

\vspace{-1.2cm}
\begin{IEEEbiography}[{\includegraphics[width=1in,height=1.25in,clip,keepaspectratio]{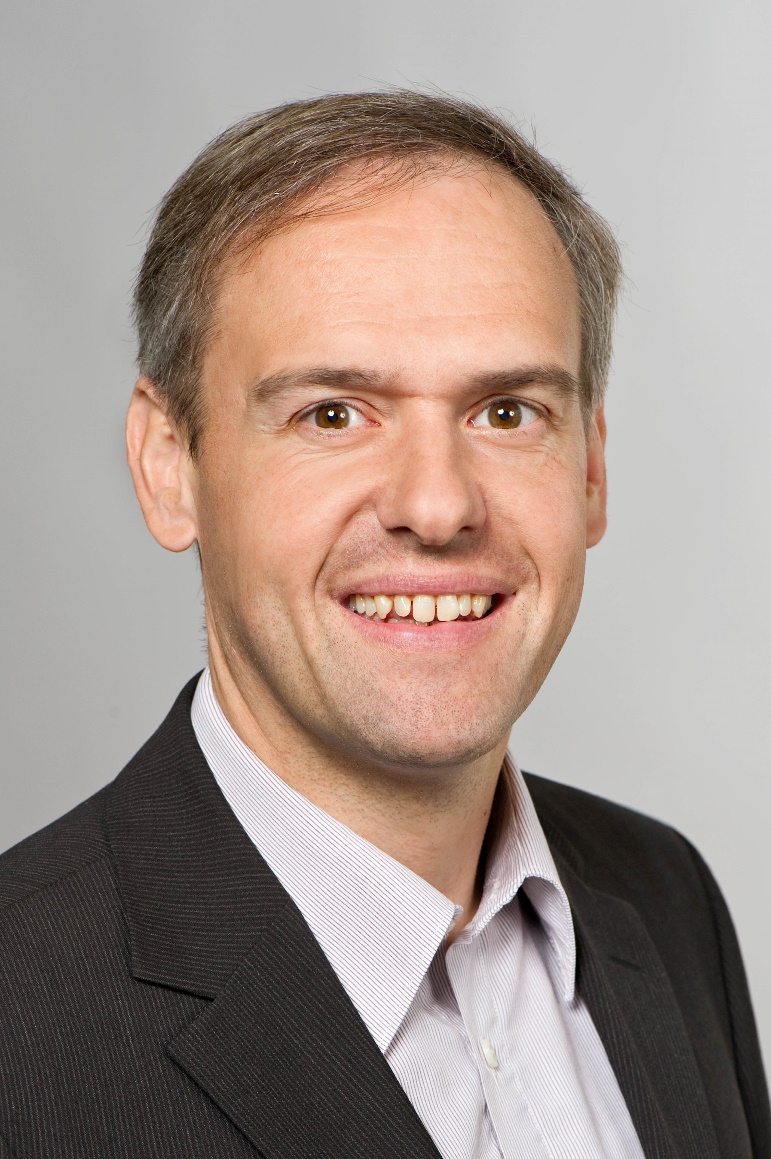}}]{Wolfgang Kellerer}
(M$'$96 \textendash\ SM$'$11) is a Full Professor with the Technical University of Munich (TUM), heading the Chair of Communication Networks at the Department of Electrical and Computer Engineering. Before, he was for over ten years with NTT DOCOMO's European Research Laboratories. He received his Dr.-Ing. degree (Ph.D.) and his Dipl.-Ing. degree (Master) from TUM, in 1995 and 2002, respectively. His research resulted in over 200 publications and 35 granted patents. He currently serves as an associate editor for IEEE Transactions on Network and Service Management and on the Editorial Board of the IEEE Communications Surveys and Tutorials. He is a member of ACM and the VDE ITG. \end{IEEEbiography}



\end{document}